\documentclass[pdflatex,sn-mathphys-num]{sn-jnl}
\usepackage{graphicx}%
\usepackage{multirow}%
\usepackage{amsmath,amssymb,amsfonts}%
\usepackage[title]{appendix}%
\usepackage{xcolor}%
\usepackage{textcomp}%
\usepackage{manyfoot}%
\usepackage{booktabs}%
\usepackage{algorithm}%
\usepackage{algorithmicx}%
\usepackage{algpseudocode}%
\usepackage{listings}%

\usepackage[caption=false]{subfig}%
\usepackage[section]{placeins}%
\usepackage[T1]{fontenc}

\newcommand{\sig}{\sigma}
\newcommand{\al}{\alpha}
\newcommand{\la}{\lambda}
\newcommand{\bol}{\boldsymbol}
\DeclareMathAlphabet\mathsfbi{OT1}{cmss}{bx}{n}
\IfFileExists{t1phv.fd}{%
  \DeclareMathAlphabet\mathsfbi{T1}{phv}{b}{it}%
}{}
\IfFileExists{ot1phv.fd}{%
  \DeclareMathAlphabet\mathsfbi{OT1}{phv}{b}{it}%
}{}

\title[Article Title]{Modelling the onset and evolution of immiscible viscous fingering in porous media}

\author*[1]{\fnm{Paulo L. K.} \sur{Caetano Chang}}\email{Paulo.Chang@student.uib.no}
\author[1]{\fnm{Kundan} \sur{Kumar}}\email{kundan.kumar@uib.no}
\author[2]{\fnm{Arne} \sur{Skauge}}\email{a.skauge@hw.ac.uk}
\author[2]{\fnm{Kenneth S.} \sur{Sorbie}}\email{k.sorbie@hw.ac.uk}

\affil[1]{\orgdiv{Department of Mathematics}, \orgname{University of Bergen}, \orgaddress{\street{Allégaten 41}, \city{Bergen}, \postcode{5007}, \state{Vestland}, \country{Norway}}}

\affil[2]{\orgdiv{School of Energy, Geoscience, Infrastructure and Society}, \orgname{Heriot-Watt University}, \orgaddress{\city{Edinburgh}, \postcode{EH14 4AS}, \country{Scotland, UK}}}

\abstract{\unboldmath%
    The simulation of viscous fingering in porous media is of direct relevance to displacement processes in petroleum engineering and hydrogeology. Building on recent work proposing a modelling approach for well-defined fingers at very adverse viscosity ratios, we investigate the physical mechanisms behind viscous fingering and the modelling requirements for capturing the finger scales and saturation patterns observed in experiments. We simulate and match a viscous fingering experiment at a viscosity ratio of $\mu_{o}/\mu_{w}{=}2000$, discussing the physical significance of each modelling step. Linear stability analysis is used to characterize the early-stage instability of the displacement. Subsequent numerical simulations show that, for the simulated finger scales to match the experiment, the most unstable wavelength at onset must be several times smaller than the desired finger width---so that, after accounting for shielding and merging in the nonlinear regime, the fingers remain thin. Small-scale channelling effects are also required to disrupt the trailing stable region commonly observed in simulations of viscous fingering in nearly homogeneous media. Finally, we show that including a weakly oil-wet capillary pressure function enables our model to capture the bypassed oil observed in the experiment.
}


\begin{document}
\maketitle

\section{Introduction}\label{sec:intro}
The simulation of viscous fingering and channelling in porous media has attracted significant amount of attention in the literature, due to its relevance to displacement processes in petroleum engineering and hydrogeology. Both fingering and channelling can occur simultaneously, and they can be difficult to disentangle—as they are not mutually exclusive, although they are different in nature. In viscous fingering, the difference in fluid properties (e.g.\ viscosity and/or density) between the displacing and displaced fluids is the driver of instability at the displacement interface, resulting in the formation of finger-like patterns. Channelling, on the other hand, refers to the preferential flow paths created by heterogeneity (of the porous medium), which can also significantly distort the displacement front. 

There is an extensive body of research in the literature on both miscible and immiscible viscous fingering in porous media. This has been generated since the pioneering early works by \citet{hill1952} and by \citet{saffman1958}. The reader is directed to reviews by \citet{homsy1987} and \citet{salmo2022}, covering the main aspects of the field. \citet{homsy1987} offers a classical review on both miscible and immiscible viscous fingering, while \citet{salmo2022} offers a more recent review of the literature on immiscible viscous fingering, which they group under the categories of linear stability analysis, direct simulation, experimental studies of unstable flow and pore-scale modelling. \citet{salmo2022} note that the vast majority of research on immiscible viscous fingering is theoretical or numerical, with only a few published experimental studies providing unstable fingering results at the Darcy scale (core floods and slab/porous-pack floods) suitable for validating continuum simulation models.  % chktex 2 chktex 12

Two research groups have published an extensive set of experimental data which can be used for modelling viscous fingering: the group of Mohanty et al.\ at the University of Texas at Austin, who performed many core floods and micromodel work \citep{doorwar2011,doorwar2014a,doorwar2014b,doorwar2017,worawutthichanyakul2017}; and Skauge et al.\ at the University of Bergen in Norway who published a number of immiscible unstable displacement in larger sandstone rock slabs \citep{skauge2009,skauge2011,skauge2012,skaugeT2014}.

In this work, we use the slab flood results published by Skauge and co-workers. These experiments consist of water displacing oil in a slab of Bentheimer sandstone, with different viscosity ratios ($\mu_o/\mu_w{=}$400, 600, 2000, and 7000). In \citet{salmo2022}, the authors were able to model the finger evolution, up to breakthrough, seen in water-displacing-oil experiments for all these viscosity ratios. They did so by using random correlated permeability fields, and relative permeability functions that sought to maximize mobility for a given fractional curve, which allowed their models to match the early-stage finger patterns and the production curves (water-cut and oil recovery). However, for the cases with $\mu_o/\mu_w{=}$400, 600 and 2000, significant areas of bypassed oil could be seen at the end of the experiments, after 1 or more pore-volumes of water injected (PVI), and these bypassed regions were not captured very well by their models.

Our main objectives in this work are (i) to develop a numerical simulation model that can capture both early and late-stage results of these viscous fingering experiments, and (ii) to provide insights on the physical mechanisms behind the observed patterns, and some guidelines for matching immiscible viscous fingering experiments. In this study we focus on the E2000 experiment, where a 1 mPa${\cdot}$s water displaces a 2000 mPa${\cdot}$s oil, in an aged sandstone slab (see Section~\ref{sec:E2000} for the details of this experiment).

Building on the work of~\citet{salmo2022}, we show that for a weakly oil-wet system, a larger-scale heterogeneity in the permeability field---superimposed on the random correlated field---can reproduce the oil bypass observed in E2000, even when the heterogeneity magnitude is relatively small. We further show that linear stability analysis can serve as a guide for selecting the flow functions (relative permeability and capillary pressure) that reproduce the degree of viscous fingering observed in the experiment. Matching the simulated finger scales to the experiment requires a system in which the most unstable wavelength is multiple times smaller than the desired finger width. Finally, we demonstrate that small-scale channelling effects are necessary to disrupt the \textit{rarefaction} zone (Section~\ref{sec:randcorr_perm}), a common feature of numerical simulations of viscous fingering in mostly homogeneous media, where a stable region trails the fingers. While a unique predictive model for such systems remains out of reach, we develop a set of physics-guided constraints that capture the observed dynamics and provide a template for modelling other high mobility-contrast multiphase displacements in porous media.

\subsection*{Relevant literature on linear stability analysis and capillary heterogeneity effects}
Linear stability analysis is a useful tool for understanding the onset of instability. Its application to viscous fingering traces back to \citet{saffman1958} and \citet{chuoke1959}. For miscible displacement, several works have followed, with notable examples including \citet{tan1986}, \citet{dewit1997}, \citet{riaz2003}, and \citet{jangir2021}. Similarly, extensive research exists for immiscible viscous fingering, including important contributions by \citet{yortsos1986}, \citet{yortsos1989}, \citet{chikhliwala1988a}, \citet{riaz2004}, and \citet{daripa2008}. \citet{caetanochang2026arxiv} extend linear stability analysis to a partially miscible displacement of a two-phase, two-component system. Finally, \citet{chikhliwala1988b} carried out a weakly nonlinear stability analysis for immiscible displacement, finding a supercritical state near the cutoff wavenumber where the perturbations stabilize at a finite amplitude. % chktex 12

To study the evolution of viscous fingers in the nonlinear regime, however, numerical simulations are usually necessary. Notable works on immiscible viscous fingering simulation include~\citet{riaz2006b} and~\citet{riaz2006a}, where the authors present numerical investigations on the nonlinear behaviour of viscous fingers and the impact of relative permeability on the fingering process. Their insights, particularly their results on the effect of relative permeability, contributed to the work of \citet{sorbie2020} and \citet{salmo2022}.
The latter two papers, developed an approach based on starting with the fractional flow and selecting the maximum mobility relative permeabilities (flow functions) in order to produce fully developed viscous fingering. As a result, \citet{salmo2022} was one of the first studies to present a good comparison between Darcy scale (continuum) numerical simulations of immiscible viscous fingering and the experimental slab flood results of \citet{skaugeT2014} for a wide range of viscosity ratios. However, their model did not capture the bypassed oil observed in the experiments, which is one of the main focuses of our work.

Another important topic for this study is the effects of capillary heterogeneity on immiscible displacements in porous media. Several studies have showed that capillary heterogeneity, induced by variations in the permeability and/or porosity of the medium, can lead to trapping (at the macroscopic, small scale) of the wetting phase, and reduced displacement efficiency when the displaced phase is preferentially wetting \citep{yortsos1990,chang1992,chaouche1994,vanduijn1995,vanduijn2007,vives1999,worawutthichanyakul2017}. \citet{chaouche1994}, for example, investigated the effect of permeability variations along the direction of flow, and found through numerical simulations and experiments that, when sharp transitions in permeability occur, a large amount of wetting phase remains trapped behind the transition to higher permeability. In \citet{vives1999}, experiments with adverse-mobility displacements showed that the displacement was significantly more unstable when the displaced phase was preferentially wetting, than when the displaced phase was non-wetting. Likewise, \citet{worawutthichanyakul2017} found in their immiscible, unstable displacement experiments, that varying the flow rate has distinct effects on recovery (at breakthrough) depending on the wettability of the phases. For a decrease in the flow rate (which raises the ratio of capillary to viscous forces) the recovery was increased when the invading phase was preferentially wetting, while the opposite was true when the wetting phase was the defending phase. A recent paper by \citet{beteta2023} presents a somewhat different interpretation of this effect of wettability and using the immiscible viscous fingering scheme proposed earlier \citep{sorbie2020}, they were able to propose a detailed explanation of why the oil-wet case is more likely to show viscous fingering, and indeed they presented some experimental demonstration of this. % chktex 2 chktex 12

\section{Theory}\label{sec:theory}
In this section we briefly describe the governing equations for a two-phase, immiscible displacement in a porous medium and the effects of capillary heterogeneity on the flow. 
The linear stability analysis (LSA) performed in Section~\ref{sec:flowfunc_lsa} follows the formulation developed by \citet{riaz2004}, and a summary of the main steps is given in the supplementary material provided with this paper \citep{supmat}. 

The governing equations for a two-phase, immiscible displacement in porous media are well known and given by:
\begin{align}
    &\phi \frac{\partial S_\al}{\partial t} + \nabla \cdot \bol{u}_\al = 0, \label{eq:matbal} \\
    &\bol{u}_\al = -k \frac{k_{r\al}}{\mu_\al} \nabla P_\al, \label{eq:darcy} \\
    &\nabla \cdot \left( \bol{u}_w + \bol{u}_o \right) = 0, \label{eq:continuity} \\
    &P_c = P_o - P_w, \label{eq:cap_pressure}
\end{align}
where $\bol{u}_\al$ is the Darcy velocity of phase $\al(=w,o)$, $\phi$ is the porosity, $k$ is the absolute permeability, $k_{r\al}$ is the relative permeability of phase $\al$, $\mu_\al$ is the viscosity of phase $\al$, $P_\al$ is the pressure of phase $\al$, and $P_c$ is the capillary pressure.

The equations can be nondimensionalized by the following transformations:
\begin{equation}\label{eq:normalization}
    \begin{aligned}
        \bol{x}^*=&\frac{\bol{x}}{L}, &\bol{u}^*=&\frac{\bol{u}}{U}, &t^*=&\frac{U}{\phi_0 L}t,\\
        \la_\al^*=&\tau k_{r\al}, &P_\al^*=&\frac{k_0}{U\mu_\al L}P_\al, &P_c^*=&\frac{k_0}{U\mu_w L}P_c 
    \end{aligned}
\end{equation}
where $U$ is the injection specific discharge (or Darcy velocity), $k_0$ is the reference permeability, and $\phi_0$ is the reference porosity, with:
\begin{equation}\label{eq:tau_varphi}
    k(\bol{x}) = \tau(\bol{x}) k_0 \quad \text{and} \quad \phi(\bol{x}) = \varphi (\bol{x}) \phi_0.
\end{equation}

Equations~\ref{eq:matbal}--\ref{eq:cap_pressure} can be rewritten as:
\begin{align}
    &\varphi \frac{\partial S_\al}{\partial t^*} + \nabla_{*} \cdot \bol{u}_\al^* = 0, \label{eq:matbal*} \\
    &\bol{u}_\al^* = \la_\al^* \nabla_{*} P_\al^*, \label{eq:darcy*} \\
    &\nabla_{*} \cdot \left( \bol{u}_w^* + \bol{u}_o^* \right) = 0, \label{eq:continuity*} \\
    &P_c^* = M P_o^* - P_w^*, \label{eq:cap_pressure*}
\end{align}
where $M = \mu_o/\mu_w$ is the viscosity ratio and $\nabla_* = \partial/\partial x_i^*$.

From Equations~\ref{eq:darcy*} and~\ref{eq:cap_pressure*}, the water phase velocity, $\bol{u}_w^*$, can be written as a function of the total velocity, ${\bol{u}_t^* = \bol{u}_w^* + \bol{u}_o^*}$, and the capillary pressure, $P_c^*$, as:
\begin{equation}\label{eq:uw}
    \bol{u}_w^* = F_w \bol{u}_t^* + \frac{\la_w^* \la_o^*}{\la_t^*} \nabla_* P_c^*,
\end{equation}
where $\la_t^* = M\la_w^* + \la_o^*$ is the total mobility and $F_w = M \la_w^* / \la_t^*$ is the fractional flow of the water phase.

\subsection{Capillary heterogeneity effects}\label{sec:capillary_heterogeneity}
As discussed above, several studies have shown that capillary heterogeneity, induced by variations in the permeability and/or porosity of the medium, can lead to trapping (at the small scale) of the wetting phase, leading to reduced displacement efficiency when the displaced phase is preferentially wetting. To see why, we consider a heterogeneous porous medium with spatially varying permeability and constant porosity (for simplicity).

We assume that the capillary pressure is a function of the saturation, porosity, and permeability of the medium, using the Leverett J-function model:
\begin{equation}\label{eq:leverett}
    P_c = \gamma \cos{\theta} \sqrt{\frac{\phi}{k}} J_c(S_w),
\end{equation}
where $\gamma$ is the interfacial tension between the two phases, $\theta$ is the contact angle, and $J_c(S_w)$ is a dimensionless function of the saturation---the Leverett J-function.

An important parameter in this context is the capillary number, which quantifies the relative importance of viscous forces to capillary forces. We define it as:
\begin{equation}
    \mathrm{Ca} = \frac{U \mu_w}{\gamma \cos\theta}.
\end{equation}

By choosing the length scale $L$ as:
\begin{equation}\label{eq:lenghtscale}
    L = \frac{\sqrt{\phi_0\, k_0}}{\mathrm{Ca}} = \frac{\sqrt{\phi_0\, k_0}}{U \mu_w} \gamma \cos\theta,
\end{equation}
we can write, using Equation~\ref{eq:tau_varphi}, the dimensionless capillary pressure as:
\begin{equation}\label{eq:leverett*}
    P_c^* = \tau^{-\tfrac{1}{2}}(\bol{x})\, J_c(S_w).
\end{equation}

Note that the presence of capillary number appears in the definition of the length scale means that Ca controls the scale $L$ where capillary pressure effects become significant. For convenience, we omit the asterisks from here on, with the understanding that all variables are now dimensionless. To observe the effects of $P_c$ on flow in a heterogeneous porous medium, we turn to the advection-diffusion equation that can be derived by substituting Equation~\ref{eq:uw} in~\ref{eq:matbal*}:
\begin{equation}\label{eq:adv_diff}
    \frac{\partial S_w}{\partial t} + \nabla \cdot \left( F_w \bol{u}_t + \frac{\la_w \la_o}{\la_t} \nabla P_c \right) = 0.
\end{equation}

By expanding the gradient of the $P_c$ in the above equation, we arrive at:
\begin{equation}\label{eq:adv_diff2}
    \frac{\partial S_w}{\partial t} + \nabla \cdot \bigg[ \underbrace{\frac{}{} F_w \bol{u}_t}_{\text{advective}} + \underbrace{\la_r \tau^{\tfrac{1}{2}} \frac{\mathrm{d} J_c}{\mathrm{d} S_w} \nabla S_w}_{\text{diffusive}} - \underbrace{\frac{\la_r}{2} \tau^{-\tfrac{1}{2}} J_c \nabla \tau}_{\text{cap-heterogeneity}} \bigg] = 0,
\end{equation}
where $\la_r = k_{rw} k_{ro} / (M k_{rw} + k_{ro})$.

The main effect of modelling $P_c$ as a function of both $S_w$ and $k$ is that the non-advective (the capillary) term is split into a diffusive term (where the driving force is $\nabla S_w$) and a capillary heterogeneity term (where the driving force is $\nabla \tau$). The diffusive term always has a stabilizing effect as it acts to reduce the saturation gradient (since by definition $\partial J_c/\partial S_w {\leq} 0$). The capillary-heterogeneity term, however, is advective: $\lambda_r$ and $J_c$ are functions of $S_w$, while $\nabla\tau$ depends on space only. Its effect can be either stabilizing or destabilizing, depending on the sign of $J_c$. If the medium is oil-wet ($J_c {<} 0$), the term tends to be destabilizing, as it pushes the water phase from low- to high-permeability regions, where preferential flow paths naturally develop; if the medium is water-wet ($J_c {>} 0$), it tends to be stabilizing, pushing the water phase from high- to low-permeability regions.

More importantly, if the displaced (oil) phase is preferentially wetting, then capillary heterogeneity term act as barrier to entry---to the invading (water) phase---into the lower permeability regions, and may explain the presence of untouched oil in the late stage of experiment E2000, as shown in Section~\ref{sec:results}. 

Finally, we can compare the relative strength between the heterogeneity and diffusive terms by looking at the ratio of their absolute values:
\begin{equation}\label{eq:het_diff_ratio}
    R_{\text{het}/\text{dif}} = \frac{1}{2} \frac{\left| J_c \right|}{\left| J_c' \right|} \frac{\left\| \nabla \tau \right\|}{\tau} \frac{1}{\left\| \nabla S_w \right\|},
\end{equation}
where $\left\| \cdot \right\|$ denotes the Euclidean norm.

So the importance of the capillary-heterogeneity term (relative to the diffusive term) increases with ${\left| J_c \right| / \left| J_c' \right|}$ and ${\left\| \nabla \tau \right\| / \tau}$. The first ratio gives us a clue as to which kinds of $P_c$ functions are more likely to have a strong heterogeneity effect, viz.\ those with relatively large absolute values and small slopes. The second ratio shows that the capillary-heterogeneity effect is more significant where the permeability changes rapidly (between highs and lows) and where the permeability is low (overall).

\subsection{Relative permeability and capillary pressure functions}\label{sec:relperm_cap_pressure}
To model the relative permeability (RP) functions we use the LET correlation \citep{lomeland2005}, which for a two-phase water-oil system can be written as:
\begin{align}
    k_{rw} &= k_{rwf} \frac{{S_{wn}}^{L_w}}{{S_{wn}}^{L_w} + E_w {\left( 1-S_{wn} \right)}^{T_w}}, \label{eq:krw_let} \\
    k_{ro} &= k_{rof} \frac{{\left( 1-S_{wn} \right)}^{L_o}}{{\left( 1-S_{wn} \right)}^{L_o} + E_o {S_{wn}}^{T_o}}, \label{eq:kro_let}
\end{align}
where
\begin{equation}
    S_{wn} = \frac{S_w - S_{wr}}{1 - S_{wr} - S_{or}},
\end{equation}
and $S_{wr}$, $S_{or}$ are the residual water and oil saturations, respectively. 

For the capillary pressure function, we model the Leverett J-function using the \textit{tangent} correlation proposed by \citet{foroughi2022}, given by:
\begin{equation}\label{eq:j_tangent}
    J_c(S_w) = A + B\tan\left( \frac{\pi}{2} - \pi {S_{wn}}^C \right).
\end{equation}

\section{Methodology}\label{sec:methodology}
In this section we explore the main elements of our methodology, including the viscous fingering experiment we are trying to match, the flow functions and LSA results that guide our choice of flow functions and the evolution of fingers in the nonlinear regime. 

First, we briefly describe the E2000 experiment ($\mu_o/\mu_w{=}2000$), the main features of the displacement observed during the water-flooding stage, and the scale of the viscous fingers, in Section~\ref{sec:E2000}. In Section~\ref{sec:flowfunc_lsa}, we select the parameters of the RP and $P_c$ functions (defined in Section~\ref{sec:relperm_cap_pressure}) used in the simulations and present the corresponding LSA results. We then investigate the nonlinear evolution of viscous fingers through numerical simulations. An analysis of finger evolution in a homogeneous medium is presented in Section~S3 of the supplementary text \citep{supmat} and summarized at the end of Section~\ref{sec:flowfunc_lsa}.

In Section~\ref{sec:randcorr_perm} we consider the heterogeneous case with randomly correlated permeability fields, examining how channelling affects viscous fingering. We assess, in particular, how permeability heterogeneity affects the trailing, stable displacement zone behind the fingers---usually seen in homogeneous media settings---and the tortuosity of the fingers.

Finally, in Section~\ref{sec:guidelines} we use the insights from Sections~\ref{sec:E2000}--\ref{sec:randcorr_perm} to create a set of guidelines for modelling a viscous fingering experiment. We then apply these guidelines in Section~\ref{sec:results} to match the results of experiment E2000. 

In the following sections, the viscosities, porosity, average permeability, and injected specific discharge of the simulation models match those of experiment E2000 (see Section~\ref{sec:E2000}). Table~\ref{tab:models} summarizes the other main characteristics of the numerical models for each particular section. Also, we report our simulation results in two interchangeable, dimensionless time scales: pore volumes injected (PVI) and the dimensionless time $t^*{=}t/t_{RP1}$, where $t_{RP1}$ is the breakthrough for the one-dimensional, Buckley-Leverett solution (for $P_c{=}0$), with RP1 relative permeability functions defined in Table~\ref{tab:relperms}. Because in all simulations the injected specific discharge, porosity and distance to the outlet are the same as in experiment E2000, PVI and $t^*$ are also comparable between the different models. For reference, here 1 PVI corresponds to approximately 7.9 days, and $t_{RP1}{=}1.23$ days, so $t^*{=}1 \approx 0.156\,$PVI.\@

\begin{table}[!htbp]
    \centering
    \caption{Summary of the models used in the simulations. Flow is in the $x$-direction.}\label{tab:models}
    \begin{tabular}{lccccc}
        \toprule
        \multirow{2}{*}{Section} & Permeability & \multirow{2}{*}{Rel. Perm./$P_c$} & Dimensions & N.\ cells \\ 
         & field &  & ($L_x{\times}L_y{\times}L_z$) & ($N_x{\times}N_y{\times}N_z$) \\ %[9pt]
        \midrule
            {\ref{sec:randcorr_perm}} & Random correlated & RP1-Pc1/RP2-Pc2 & $30{\times}30{\times}2.55\,$cm & $1000{\times}1000{\times}1$ \\
            \multirow{2}{*}{\ref{sec:results}} & Combination of two & \multirow{2}{*}{RP2-Pc2} & \multirow{2}{*}{$30{\times}30{\times}2.55\,$cm} & \multirow{2}{*}{$1000{\times}1000{\times}1$} \\
            & correlation lengths & & & & \\
        \bottomrule
    \end{tabular}
\end{table}

All simulations are performed using the commercial black-oil simulator IMEX, developed by Computer Modelling Group (CMG) \citep{imx2025}.

\FloatBarrier\subsection{The E2000 experiment}\label{sec:E2000}
\begin{figure}[!htbp]
    \centering
    \includegraphics[width=\textwidth]{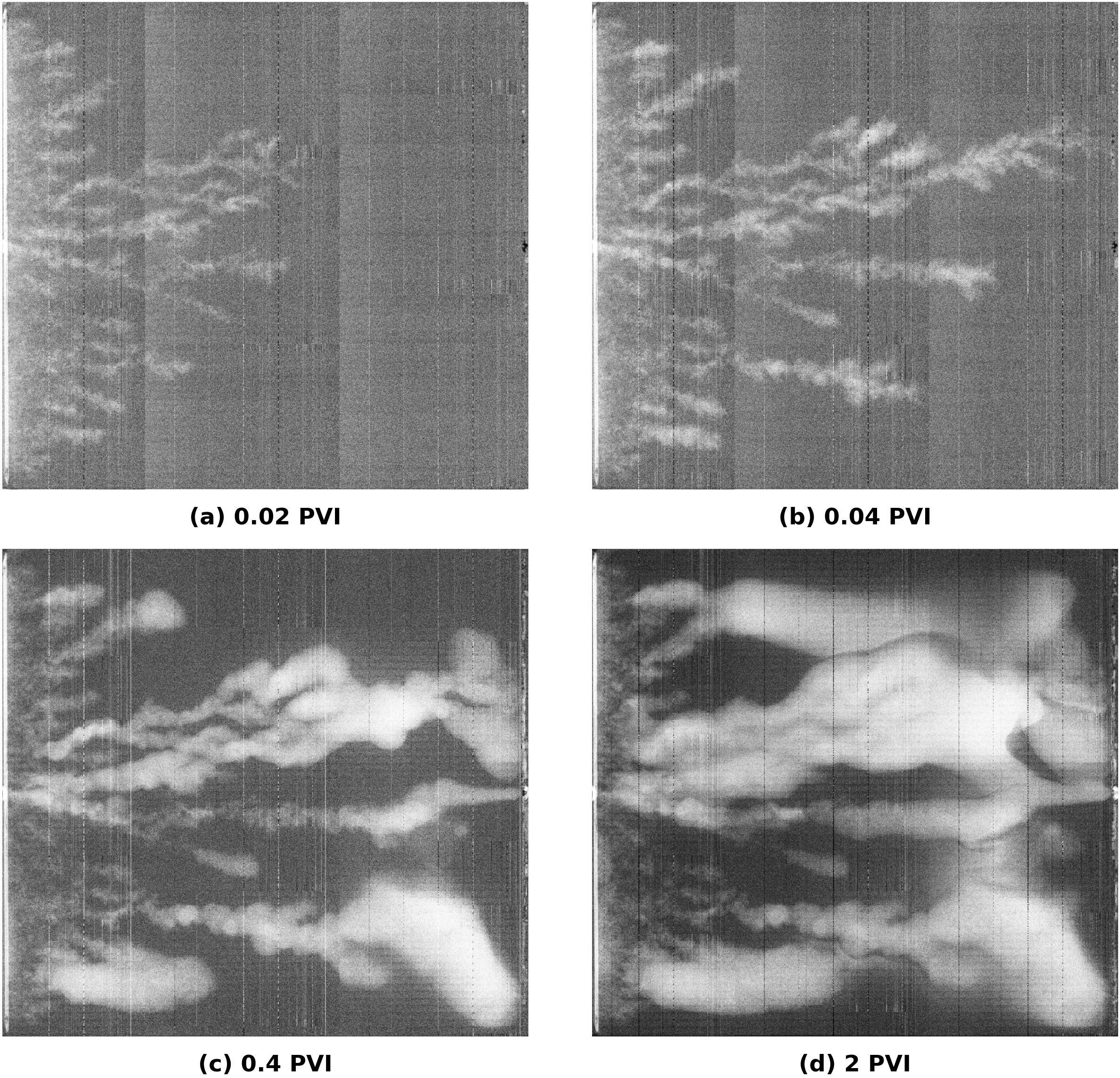}
    \caption{X-ray pictures of the experiment with a 2000 viscosity ratio at different moments, given in pore-volumes injected (PVI): (a) early-stage (0.024 PVI); (b) breakthrough (0.04 PVI); (c) intermediate-stage (0.4 PVI); (d) late-stage (2 PVI).}\label{fig:xray_exp2000}
\end{figure}

The E2000 experiment, described in \citep{skaugeT2014}, consists of a water flood followed by a polymer flood, though our focus is on the water-flooding portion. The experiment consisted in displacing a very viscous oil ($\mu_o{=}2000\,\text{mPa}{\cdot}\text{s}$) with water ($\mu_w{=}1\,\text{mPa}{\cdot}\text{s}$) in an aged Bentheimer sandstone slab with dimensions of $30 \times 29.8 \times 2.55\,$cm, with a porosity of 24.8\% and average absolute permeability of approximately 2.5 Darcy. The specific discharge at the inlet is $U{=}6.5{\times}10^{-4}\,$cm/min. Visual inspection suggests the rock slab is mostly homogeneous, although some degree of heterogeneity is likely to be present. The ageing process is expected to have left the rock slab slightly oil-wet. Figure~\ref{fig:xray_exp2000} shows x-ray pictures of the slab at different moments during the experiment. The saturation image is given in greyscale (from 0 to 255), where the darker regions are more oil-saturated and the lighter regions are more water-saturated.\footnote{The actual saturation values were not calculated from these images, but from more accurate measurements (see Section~\ref{sec:results}). Because these measurements were more time-consuming, no such measurements were made before water breakthrough occurred.}

Four main observations can be made from Figure~\ref{fig:xray_exp2000}: 
\begin{enumerate}
    \item~The displacement is highly unstable, with many fingers developing early on, and many instances of fingers splitting;
    \item~No stable displacement region trailing the fingers is observed;
    \item~After breakthrough, the fingers grow mostly sideways and merge;
    \item~Even after 2 PVI, a significant amount of oil has been bypassed and left untouched by the injected water phase.
\end{enumerate}

Observation 2 indicates that small-scale channelling is present which, coupled with the high viscous instability, prevents the development of a stable, rarefaction zone behind the fingers (commonly observed in viscous fingering simulations in mostly homogeneous media). Section~\ref{sec:randcorr_perm} shows that random correlated permeability fields with enough variation can indeed disrupt this rarefaction zone.

Observations 3 and 4 are likely due to the presence of a larger scale of heterogeneity in the permeability field that steers finger development: after breakthrough the fingers grow sideways and merge, creating highly preferential pathways, which remain relatively static due to capillary-heterogeneity effects, leading to significant oil bypass.

Visual inspection of the rock slab used in the experiment indicated that it was mostly homogeneous, which does not necessarily preclude the presence of heterogeneity in the permeability, but it should limit the degree of contrast in the permeability field. We can reasonably assume that very small-scale variations in the permeability would remain undetected by visual inspection. For the larger scale, we assume that modest variations would also remain undetected. What is considered modest is debatable, but we assume that a distribution with ${K_{max}/K_{min}\approx3}$, where $K_{max}$ and $K_{min}$ are the maximum and minimum permeability values, is a modest enough range. Together with an oil-wet Pc, we show that this permeability contrast explains the presence of untouched oil at the late stage of the experiment.

Next, we turn to the RP and $P_c$ functions, which are the other two main elements of our model, defining the stability of the water-oil displacement.

\subsection{Flow functions and linear stability analysis}\label{sec:flowfunc_lsa}
\begin{figure}[!htbp]
    \centering
    \subfloat[PVI=0.024\label{fig:fourier_imsat_t1}]{
        \includegraphics[width=0.48\textwidth]{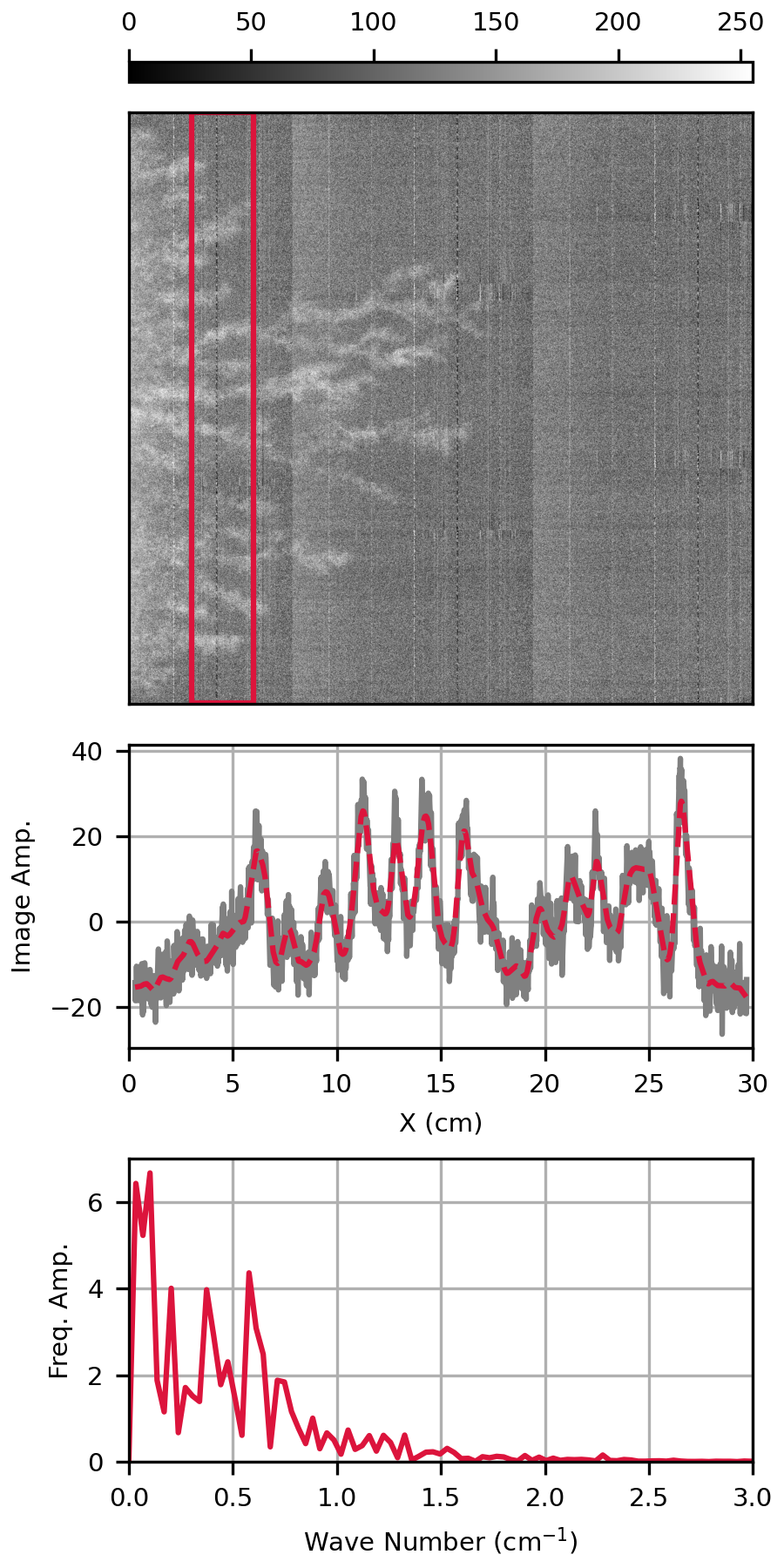}
    }\hfill
    \subfloat[PVI=0.04\label{fig:fourier_imsat_t2}]{
        \includegraphics[width=0.48\textwidth]{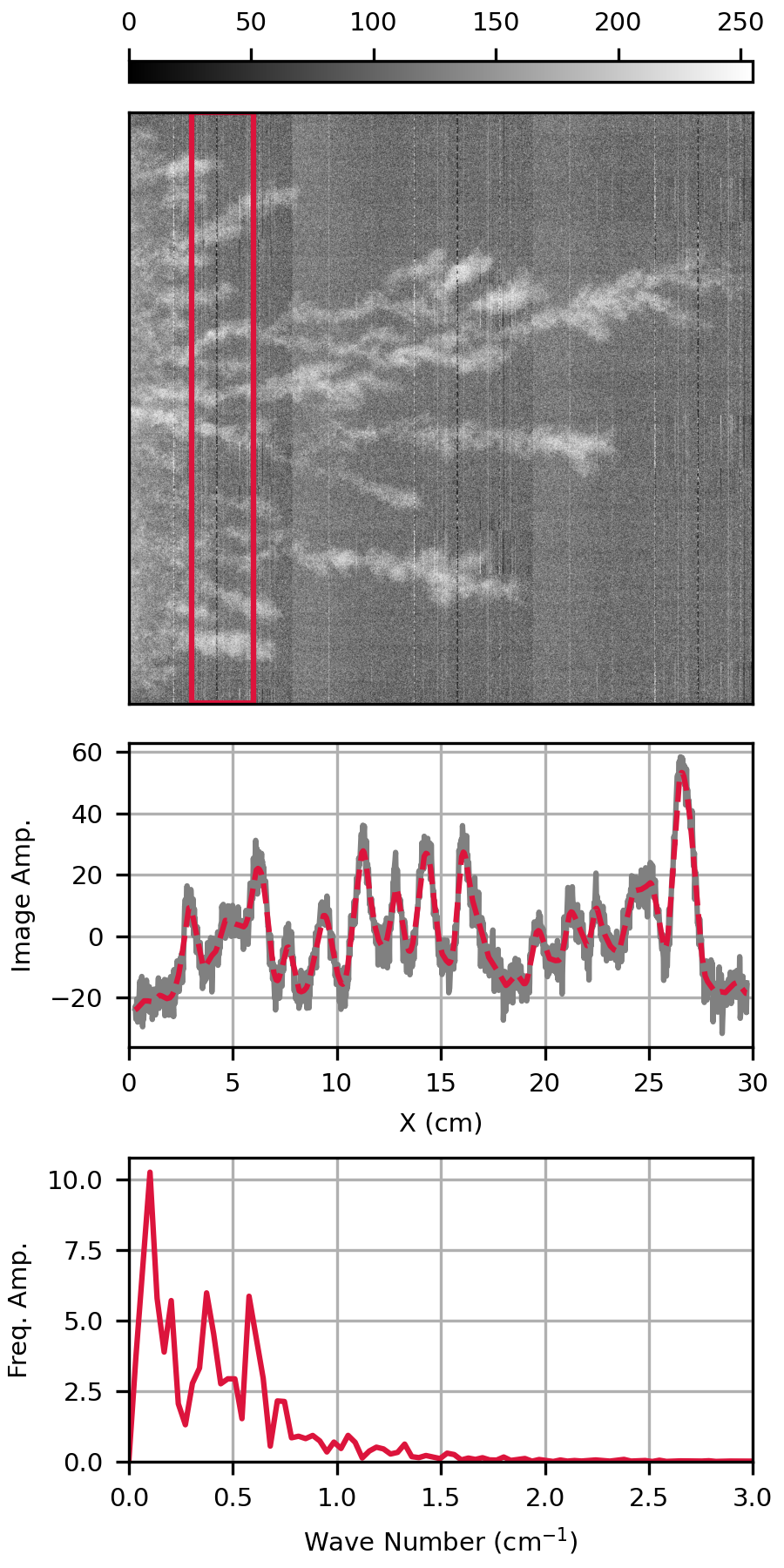}
    }
    \caption{X-ray image of the saturation (in greyscale), longitudinally averaged image (signal) and the fast Fourier transform of the smoothed signal, at two different times: (a) PVI=0.024 and (b) PVI=0.04. The red rectangle at the image indicates the analysed region and the dashed-red line shows the smoothed signal.}\label{fig:fourier_imsat}
\end{figure}

To estimate how unstable the displacement in E2000 is, we analyse the fingers seen in Figures~\ref{fig:xray_exp2000}a and~\ref{fig:xray_exp2000}b, and the results are shown in Figure~\ref{fig:fourier_imsat}. The top pictures in Figures~\ref{fig:fourier_imsat}a and~\ref{fig:fourier_imsat}b show the analysed portions of the x-ray images, defined by the red rectangles; the x-ray/saturation values inside this strip are averaged in the direction of flow, leading to a one-dimensional signal (middle picture), with its smoothed version shown as the superimposed dashed-red line. Finally, the bottom picture shows the fast Fourier transform of the smoothed signal. In both instants, PVI=0.02 and PVI=0.04, the saturation signals are similar, and so are their spectra. From the middle picture, we find that a representative value for the wavelength of the fingers is around 1.7 cm, and therefore we estimate that the dominant wavenumber\footnote{Details on the definition and calculation of the dominant wavenumber can be found in Section~S2 of the supplementary text \citep{supmat}} to be ${\hat{\nu}\approx0.6\,\text{cm}^{-1}}$. The frequency spectrum shows almost no energy at wavenumbers larger than 1 cm$^{-1}$.

It logically follows that our choice of RP and $P_c$ functions should lead to a dispersion relation with a cut-off wavenumber $\nu_{cut}$ larger than ${\thicksim 1\,\text{cm}^{-1}}$---otherwise the observed fingers would not be able to develop.
In fact, the cut-off wavenumber should be much larger than {1 cm$^{-1}$}, since shielding and merging effects means that, as the displacement progresses, the fingers that come to dominate flow are on the lower end of the wavenumber spectrum. We return to this point at the end of this section.

For our study of instability and viscous fingering simulations in the next sections, we use two pairs of RP curves. For our initial choice of RP curves (RP1), we take $S_{ws}$ and $F_{ws}$ from the history-matching results of \citet{salmo2022} for the E2000 experiment. We also select a second pair of curves (RP2), with larger values of $S_{ws}$ and $F_{ws}$; replacing RP1 for RP2 increases the instability of the displacement while keeping the Buckley-Leverett recovery approximately unchanged. The parameters and properties of RP1 and RP2 are summarized in Table~\ref{tab:relperms}.

\begin{figure}[!htbp]
    \centering
    \includegraphics[width=0.9\textwidth]{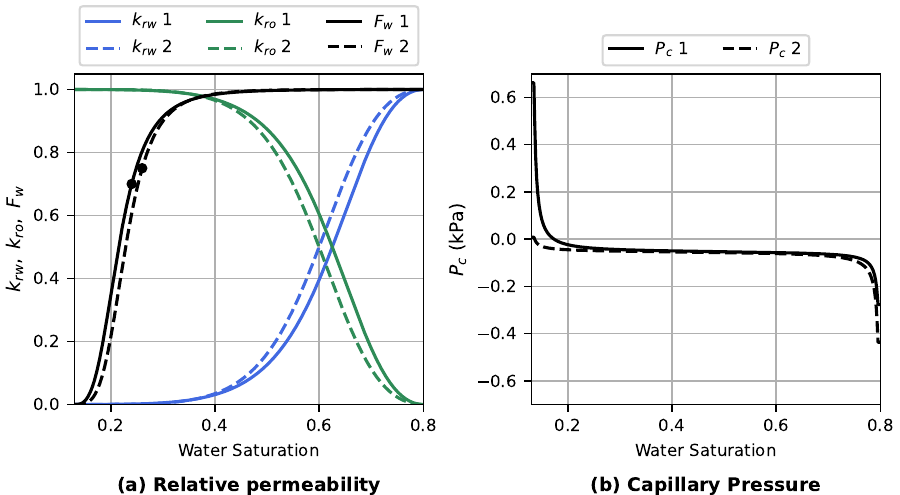}
    \caption{RP and $P_c$ functions used in the simulations. The fractional flow curve assumed $M=2000$, and the $P_c$ curve assumed $\phi{=}0.248$ and $k{=}2500\,$mD.}\label{fig:relperms_pcs}
\end{figure}

Likewise, we use two $P_c$ functions. The first, PC1, is based on matched Bentheimer sandstone data reported by \citet{foroughi2022} (see their Figure~2b\footnote{We note that the authors refer to this $P_c$ as mixed-wet, but in this study we refer to it as weakly oil-wet, since the J-function is negative for most of the saturation range.}). We selected PC1 because it is based on data from the same geological formation as E2000 and because it is weakly oil-wet, consistent with the ageing process used in the experiment. Additionally, PC1 has a relatively small slope ($|J_c'|$) near $S_{ws}$, permitting the onset of instability at large wavenumbers, while its magnitude near $S_{ws}$ ($|J_c|$) is sufficient to produce a significant capillary-heterogeneity effect. The second $P_c$ function is a modification of PC1 that increases instability by reducing the capillary pressure derivative near the shock for low $S_w$ values (achieved by reducing the value of $C$ in Equation\ref{eq:j_tangent}). Table~\ref{tab:pcs} summarizes the parameters of both $P_c$ functions, while Figure~\ref{fig:relperms_pcs} shows the RP and $P_c$ curves.

\begin{table}[!htbp]
    \centering
    \caption{LET model parameters of the RP functions. The RP of the oil phase is determined by $k_{ro}{=}1{-}k_{rw}$, so only the water phase parameters are specified. The values for the saturation and fractional flow at the shock ($S_{ws}$ and $F_{ws}$) and recovery assume $M=2000$.}\label{tab:relperms}
    \begin{tabular}{lcccccccccc}
        \toprule
        Name & $k_{rwf}$ & $L_w$ & $E_w$ & $T_w$ & $S_{wr}$ & $S_{or}$ & $S_{ws}$ & $F_{ws}$ & $V_s$ & \begin{tabular}{@{} c @{}}Recovery\\{}$(F_w{=}0.95)$ \end{tabular} \\ %[9pt] % @{} deletes the extra space before and after the content of the cell
        \midrule
        RP1 & 1.0 & 2.94 & 6.01 & 2.0 & 0.13 & 0.2 & 0.24 & 0.70 & 6.4 & 0.29\\
        RP2 & 1.0 & 3.52 & 3.20 & 2.0 & 0.13 & 0.2 & 0.26 & 0.75 & 5.8 & 0.29\\
        \bottomrule
    \end{tabular}
\end{table}

\begin{table}[!htbp]
    \centering
    \caption{Model parameters of the $P_c$ functions.}\label{tab:pcs}
    \begin{tabular}{lcccc}
        \toprule
        Name & $A$ & $B$ & $C$ & $\gamma\cos{\theta}$ \\ %[3pt]
        \midrule
        Pc1 & -0.017 & 0.002 & 1.2 & 10 mN/m \\
        Pc2 & -0.017 & 0.002 & 0.7 & 10 mN/m \\
        \bottomrule
    \end{tabular}
\end{table}

\begin{figure}[!htbp]
    \centering
    \includegraphics[width=0.8\textwidth]{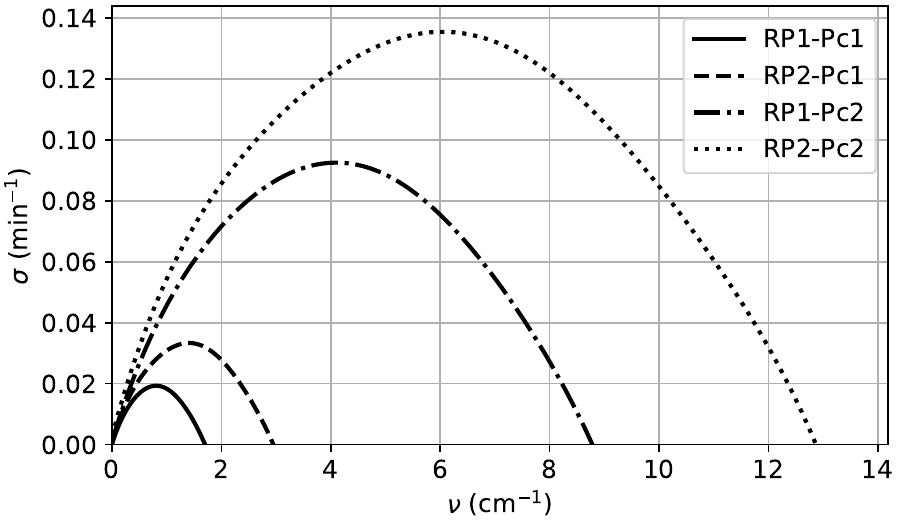}
    \caption{Dispersion relations for different combinations of RP and $P_c$ functions.}\label{fig:disprelations}
\end{figure}

In Figure~\ref{fig:disprelations}, we plot the dispersion relations---obtained by solving the eigenvalue problem of the LSA---for the different combinations of the RP and $P_c$ functions. 
These results are validated with numerical simulations in Section~S3 in the supplementary text \citep{supmat}.
In Table~\ref{tab:linstab} we list the values of the most unstable wavenumber $\nu_{max}$, its corresponding growth rate $\sig_{max}$ and the cut-off wavenumber $\nu_{cut}$ (above which all perturbations are stable) for all cases.

The first case (RP1-Pc1) has a cut-off wavenumber of $\nu_{cut}\approx1.7\,$cm$^{-1}$. Although this exceeds the amplitude spectrum of E2000 shown in Figure~\ref{fig:fourier_imsat}, we show in the next section that it is insufficient to reproduce the finger scales observed in the experiment, and that RP2-Pc2---with a $\nu_{max}$ approximately 10 times the dominant wavenumber $\hat{\nu}$ of E2000---yields fingers much closer in scale to those observed.

\begin{table}[!htbp]
    \centering
    \caption{Linear stability analysis results.}\label{tab:linstab}
    \def~{\hphantom{0}} % effect: ~ becomes a blank the width of one digit
    \begin{tabular}{lcccc}
        \toprule
        Name & $\nu_{max}$ (cm$^{-1}$) & $\sig_{max}$ (min$^{-1}$) & $\nu_{cut}$ (cm$^{-1}$) \\ %[3pt]
        \midrule
        RP1-Pc1 & 0.81 & 0.019 & ~1.70 \\ % chktex 39
        RP2-Pc1 & 1.41 & 0.033 & ~2.95 \\ % chktex 39
        RP1-Pc2 & 4.09 & 0.092 & ~8.78 \\ % chktex 39
        RP2-Pc2 & 6.04 & 0.14~ & 12.9~ \\ % chktex 39
        \bottomrule
    \end{tabular}
\end{table}

This large gap between $\nu_{max}$ and $\hat{\nu}$ is explained by the nonlinear evolution of the fingers: shielding, where larger fingers impede the growth of their smaller neighbours, and merging, where larger fingers subsequently absorb the smaller ones, leading to progressively thicker and fewer fingers as the displacement advances. These effects are discussed in detail by \citet{riaz2006b} and \citet{kampitsis2021}.

In Section~S3 of the supplementary text \citep{supmat}, we present an additional analysis of the nonlinear evolution of the fingers, showing that wavelike perturbations stabilize at finite amplitudes, as predicted by the weakly nonlinear stability analysis of \citet{chikhliwala1988b}. Although their analysis was limited to wavenumbers near the cut-off, our results suggest that this nonlinear stabilization applies to all wavenumbers, and that the final length of the fingers follows a power-law relation with their wavelengths, with an exponent of approximately 2.5.

We attribute this stabilization to the elongation of the fingers: as fingers grow longer, the transverse dispersive flow driven by capillary forces increases, eventually arresting their longitudinal growth. Importantly, this mechanism implies that thin fingers stabilize at shorter lengths than thicker ones, so that widening through merging is necessary for further growth---not shielding alone.

We can therefore expect the dominant mode of the fingers in our correlated random permeability simulations to lie on the lower end of the wavenumber spectrum by breakthrough time. That means estimating ${\nu_{max}\approx0.6\,\text{cm}^{-1}}$ from Figure~\ref{fig:fourier_imsat} is too conservative, and that when choosing the RP and $P_c$ functions we should aim for a $\nu_{max}$ well above ${0.6\,\text{cm}^{-1}}$.

\FloatBarrier\subsection{Channelling effects due to small-scale heterogeneity}\label{sec:randcorr_perm}
\begin{figure}[!htbp]
    \centering
    \includegraphics[width=\textwidth]{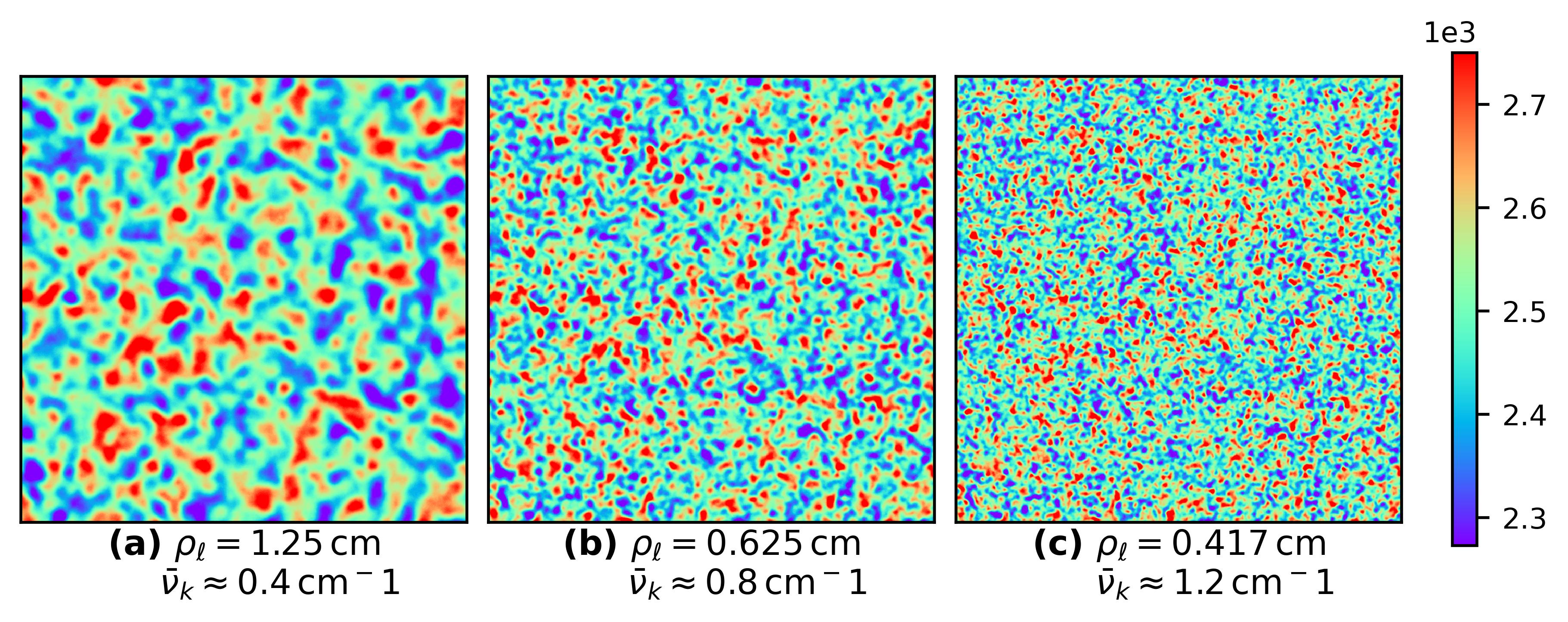}
    \caption{Permeability (in mD) fields with $\bar{\nu}_k{=}$ 0.4, 0.8 and 1.2$\,$cm$^{-1}$. For the three cases, ${k_{P50}{=}2500\,}$mD and ${k_{max}/k_{P50}{=}1.1}$.}\label{fig:corrlen_perms}
\end{figure}

In this section, we investigate viscous fingering in heterogeneous permeability fields. The grid dimensions are $30{\times}30{\times}2.55\,$cm, with 1000${\times}$1000${\times}$1 cells ($L_x{=}L_y{=}0.03\,$cm). Figure~\ref{fig:corrlen_perms} shows the permeability fields generated via Random Gaussian Simulations (RGS) with correlation lengths ($\rho_l$) of 1.25, 0.62 and 0.42$\,$cm. Assuming that the average distance between two neighbouring peaks in the permeability is approximately $2\rho_l$, we can define an ``average wavenumber'' for the permeability:
\begin{equation}
    \bar{\nu}_k = \frac{1}{2\rho_l},
\end{equation}
and the corresponding values of $\bar{\nu}_k$ for the fields in Figure~\ref{fig:corrlen_perms} are approximately 0.4, 0.8 and 1.2$\,$cm$^{-1}$. For all cases, the permeability is sampled from a log-normal distribution with a median of $k_{P50}=2500\,$mD, and the min/max values set at 2 times the standard deviation of $\ln(k)$, $\sig_{\ln}$, so that $k_{max}/k_{P50}{=}\exp(2\sig_{\ln})$.   

\begin{figure}[!htbp]
    \centering
    \includegraphics[width=\textwidth]{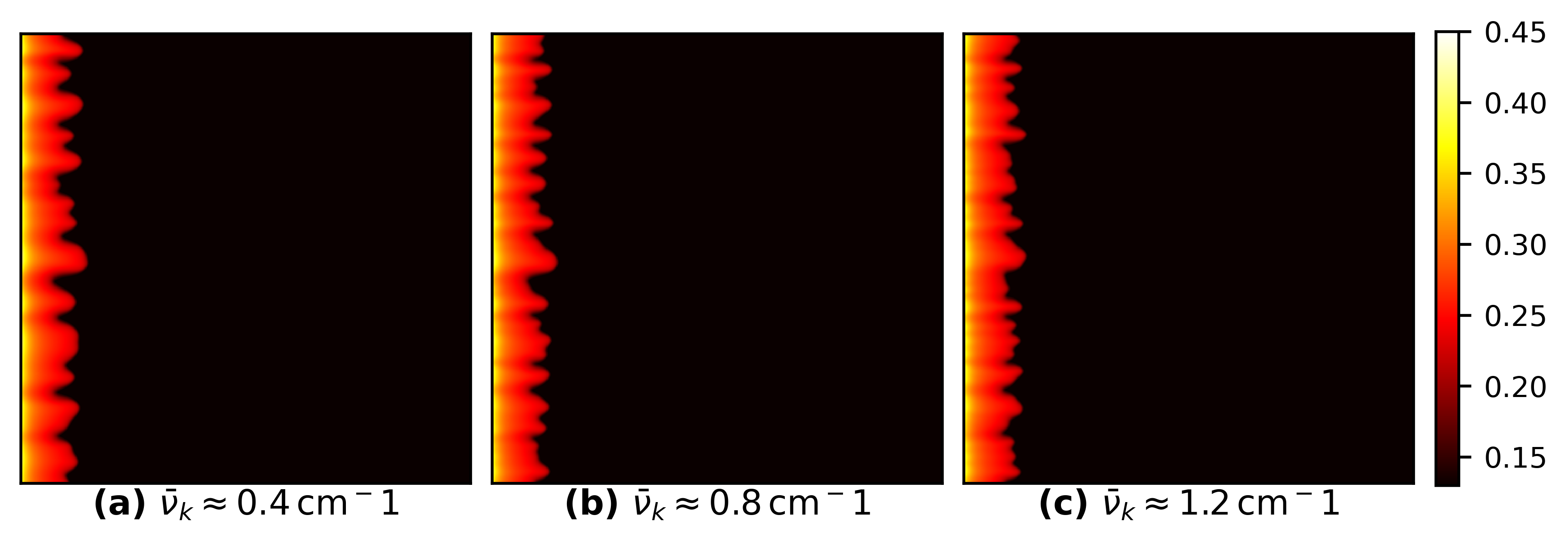}
    \caption{Saturation fields at $t^*{=}0.1$ for the cases with $\bar{\nu}_k{=}$0.4, 0.8 and 1.2 cm$^{-1}$. All fields have $k_{max}/k_{P50}{=}1.1$, and the flow functions are RP1 and Pc1. Capillary homogeneity is assumed.}\label{fig:rand_sat_rp1pc1_180}
\end{figure}

From the discussion in the previous section, we can expect that the dominant mode of the fingers will continuously decrease such that by breakthrough time the dominant wavenumber is on the lower end of the dispersion relation curve---i.e., we observe thicker fingers, and in lesser numbers, than $\nu_{max}$ from the LSA would suggest. To confirm this, we simulate the displacement for the three permeability fields shown in Figure~\ref{fig:corrlen_perms}, using the flow functions RP1 and Pc1. For this analysis, we assume capillary homogeneity, i.e., $P_c$ is a function of only the saturation (fixing $k=2.5\,$D in Equation~\ref{eq:leverett}). Figure~\ref{fig:rand_sat_rp1pc1_180} shows the saturation maps at $t^*{=}0.1$. From visual inspection, we see that the number of the fingers shows a positive correlation with the permeability field wavenumber, $\bar{\nu}_k$. At $t^*{=}0.5$ (Figure~\ref{fig:rand_sat_rp1pc1_180}), however, the fingers have merged and thickened, and the pictures for the three permeability fields look very similar. This can also be seen in Figure~\ref{fig:dominant_mode_randcorr}a, where we see that the dominant mode for all three cases decrease over time, converging to the same values at $t^*{=}0.2$. 

The more evident solution for matching the wavenumber of fingers in the nonlinear regime (as in Figure~\ref{fig:fourier_imsat}) is to select flow functions that lead to a much larger value for $\nu_{max}$ than the observed wavenumber, so that, even after the dominant mode decreases due to shielding and merging effects, the observed fingers still have a wavenumber that is consistent with the experiment. This is what we do with the RP2-Pc2 flow functions, which leads to a $\nu_{max}$ of around 6 cm$^{-1}$, around 10 times the dominant wavenumber estimated for E2000, from Figure~\ref{fig:fourier_imsat}. Figure~\ref{fig:dominant_mode_randcorr}b shows that the behaviour of the dominant modes of the RP2-Pc2 cases is similar to those of the RP1-Pc1 cases, but shifted to higher values. We confirm that in Figure~\ref{fig:rand_sat_rp2pc2_900}. An important observation is that the values of $\hat{\nu}$ does not scale linearly with $\nu_{max}$ (from the LSA)---while the RP2-Pc2 has a $\nu_{max}$ that is around 7.5 times larger than the RP1-Pc1 combination, the dominant mode throughout the simulation is only around 2 times larger.

\begin{figure}[!htbp]
    \centering
    \includegraphics[width=\textwidth]{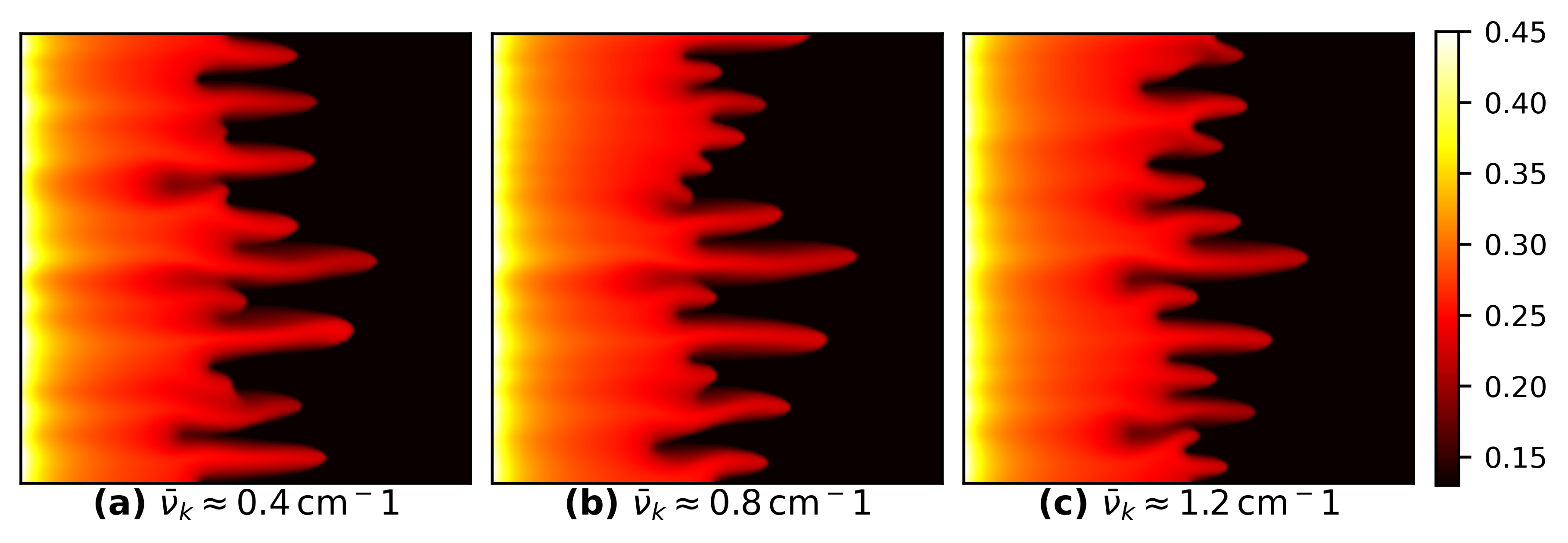}
    \caption{Same as Figure~\ref{fig:rand_sat_rp1pc1_180}, for $t^*{=}0.5$.}\label{fig:rand_sat_rp1pc1_900}
\end{figure}

\begin{figure}[!htbp]
    \centering
    \subfloat[RP1/Pc1\label{fig:dominant_mode_randcorr_rp1pc1}]{
        \includegraphics[width=0.45\textwidth]{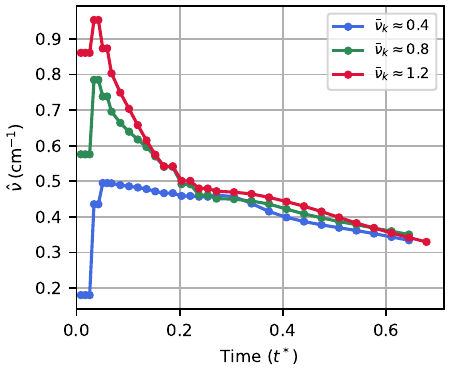}
    }\hfill
    \subfloat[RP2/Pc2\label{fig:dominant_mode_randcorr_rp2pc2}]{
        \includegraphics[width=0.45\textwidth]{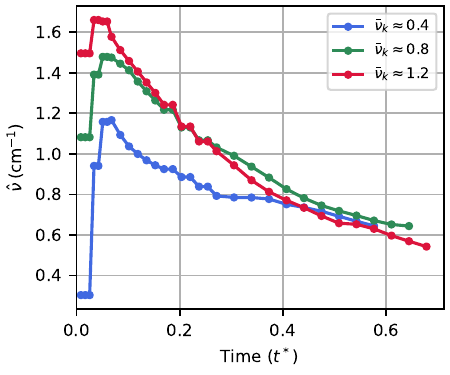}
    }
    \caption{Dominant wavenumber evolution for the random correlated permeability fields. In (a) the simulations with RP1/Pc1, and in (b) the simulations with RP2/Pc2. Capillary homogeneity is assumed.}\label{fig:dominant_mode_randcorr}
\end{figure}

\begin{figure}[!htbp]
    \centering
    \includegraphics[width=\textwidth]{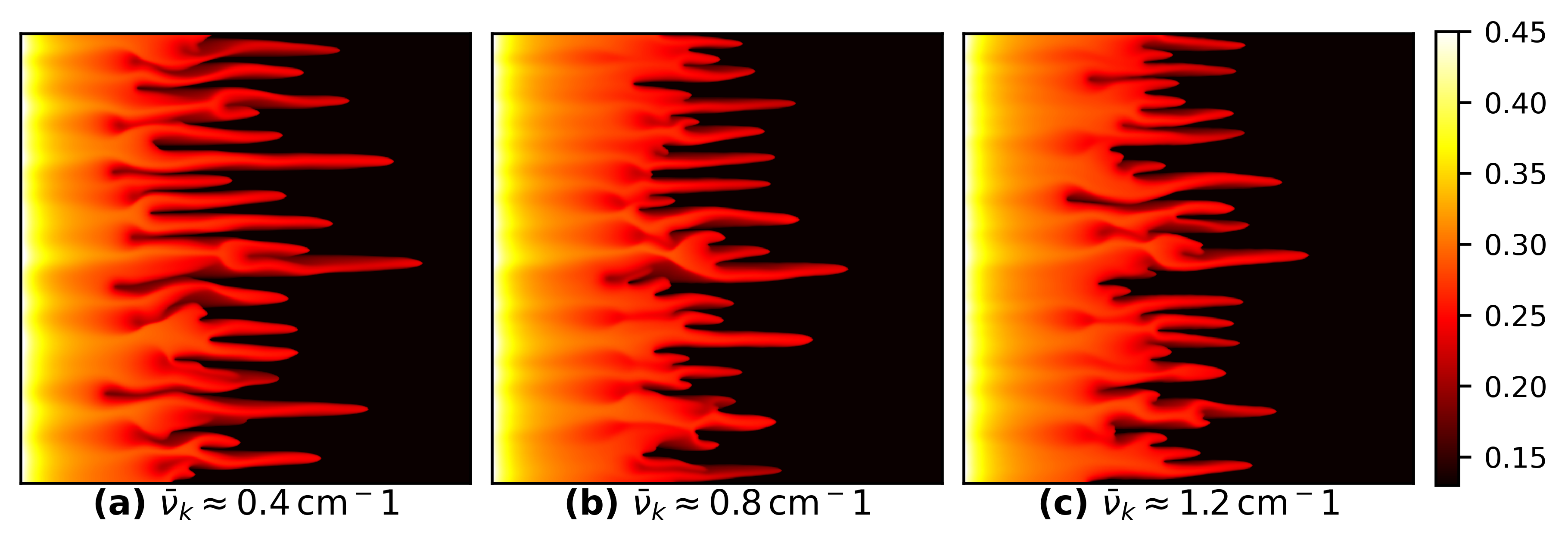}
    \caption{Same as Figure~\ref{fig:rand_sat_rp1pc1_900}, with flow functions RP2 and Pc2.}\label{fig:rand_sat_rp2pc2_900}
\end{figure}

Although the fingers for the RP2-Pc2 cases are closer in scale to the ones observed in E2000, we still see the presence of a stable zone trailing the fingers. This trailing, unperturbed zone---or \textit{rarefaction} zone---consists in the rarefaction wave that follows the Buckley-Leverett shock, for which the displacement is usually stable, at least at the scale of the shock-front instability. The presence of this zone is a common feature of numerical simulations of immiscible viscous fingering, in large part because the shock-front velocity tends to be much greater than the injection velocity---by the time breakthrough occurs, the rarefaction zone has travelled a significant fraction of the domain \citep{riaz2006b}. 

\begin{figure}[!htbp]
    \centering
    \includegraphics[width=\textwidth]{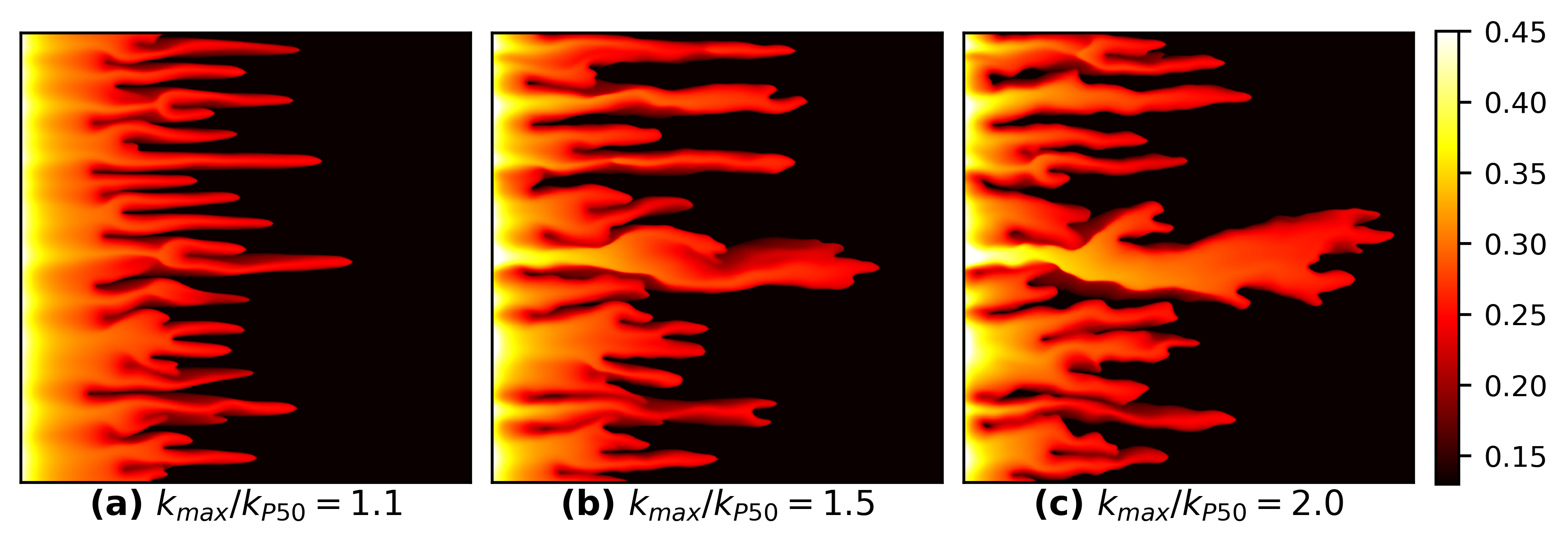}
    \caption{Saturation fields at $t^*{=}0.4$ for the cases with $k_{max}/k_{P50}{=}1.1$, 1.5 and 2. All fields have $\bar{\nu}_k{=}0.8$, and the flow functions are RP2 and Pc2. Capillary homogeneity is assumed.}\label{fig:rand_sat_deltavar_pcfix}
\end{figure}

In the E2000 experiment, however, no such zone is observed, and the fingers extend all the way to the inlet. This absence can be explained by the presence of small-scale channelling in the permeability field, which prevents the development of the rarefaction zone trailing the fingers. We test this hypothesis in Figure~\ref{fig:rand_sat_deltavar_pcfix}, where we compare saturations at $t^*{=}0.4$ for the RP2-Pc2 and $\bar{\nu}_k{=}0.4$ case, for increasing values of permeability variance. We see that, as $k_{max}/k_{P50}$ increases, the rarefaction zone is disrupted, with fingers extending all the way to the inlet; also, more instances of tip splitting occur, and the fingers become more tortuous, consistent with the E2000 observations. However, this tortuosity promotes finger coalescence, resulting in thicker fingers.

\begin{figure}[!htbp]
    \centering
    \includegraphics[width=\textwidth]{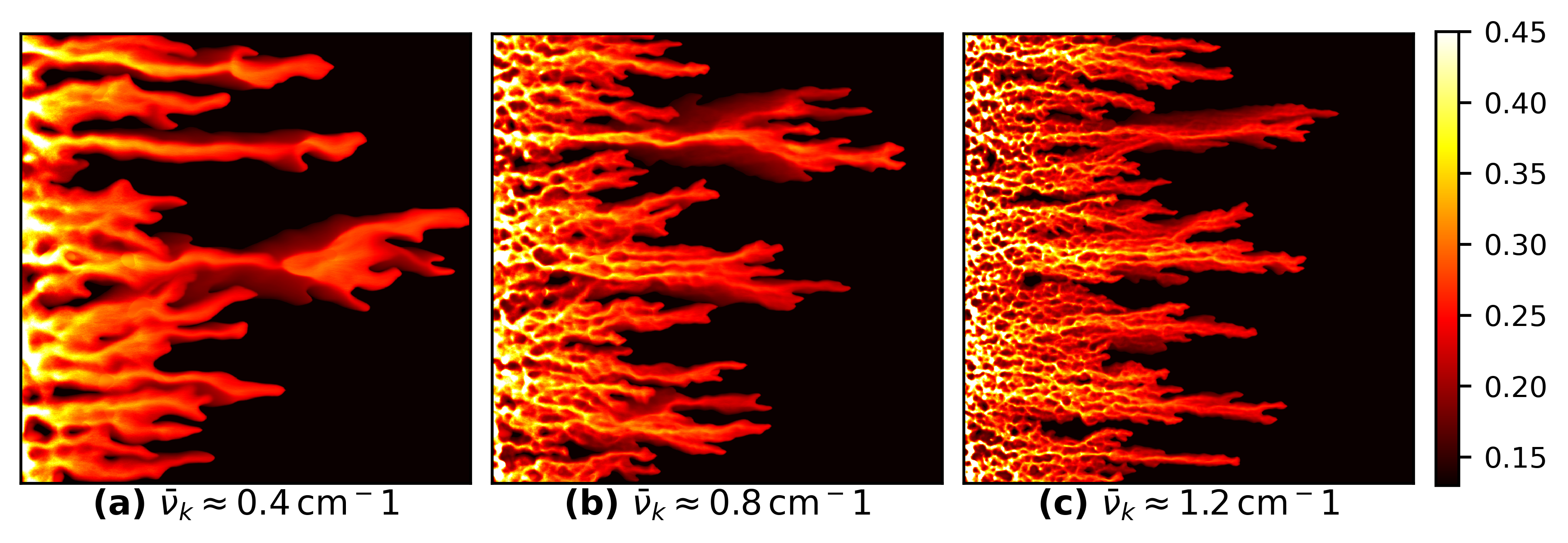}
    \caption{Saturation fields at $t^*{=}0.4$ for the cases with $\bar{\nu}_k{=}$0.4, 0.8 and 1.2 cm$^{-1}$. All fields have $k_{max}/k_{P50}{=}2$, and the flow functions are RP2 and Pc2. Capillary heterogeneity is considered.}\label{fig:rand_sat_wxvar_pcow}
\end{figure}

 So far we have only considered the capillary homogeneous case. When we consider the case with $P_c$ also depends on the permeability, a capillary-heterogeneity flow arises (Equation~\ref{eq:adv_diff2}), its magnitude proportional to the permeability gradient---in turn proportional to $k_{max}/k_{P50}$ and $\bar{\nu}_k$. In Figure~\ref{fig:rand_sat_wxvar_pcow} we plot the saturation maps of the cases using the oil-wet, permeability-dependent Pc2 function, for $k_{max}/k_{P50}{=}2$ and $\bar{\nu}_k{=}0.4$, 0.8 and 1.2 cm$^{-1}$. Comparing Figures~\ref{fig:rand_sat_deltavar_pcfix}c and~\ref{fig:rand_sat_wxvar_pcow}a, we see that the presence of capillary heterogeneity concentrates the water phase in the high permeability regions. As $\bar{\nu}_k$ increases, this effect becomes more pronounced, giving the saturation maps a ``speckled'' look (Figures~\ref{fig:rand_sat_wxvar_pcow}b-c). Such a speckled pattern has been observed in other viscous fingering experiments, e.g.\ the silanized rock slab shown in Figure 9 of \citet{beteta2024b}. This pattern is not observed, however, in the E2000 experiment.
 
There is, therefore, a balance to be struck in these parameters: the observed results appear to require (i) enough variation in the permeability field to disrupt the rarefaction zone and promote tip splitting, but (ii) not so much that the saturation field becomes too ``speckled'' by capillary effects or the fingers become too thick.

\FloatBarrier\subsection{Modelling guidelines}\label{sec:guidelines}
With the insights from the previous sections, we can now establish some guidelines for modelling not only the E2000 experiment, but also other unstable, immiscible displacements. 

First, we note that one major constraint in the choices for the RP and $P_c$ functions and permeability field (our three modelling parameters) is the production data of the experiment. In particular, the RP functions play a significant role in determining the recovery and water breakthrough time, a fact that limits the range of RP functions we can use when ``designing'' the instability of the displacement in our simulations.

For example, \citet{riaz2006a} found that a key factor to determine viscous bypass---i.e., how large fingers grow relative to the rarefaction zone---is the \textit{growth number} ($\Lambda$) of the displacement, which they define as $\Lambda{=}\sig_{max}/\nu_{max}V_s$. In particular, we found (for the flow conditions in our simulations) that increasing the shock saturation, $S_{ws}$, while keeping $F_{ws}$ constant, increases instability and reduces shock velocity $V_s$, which helps to promote viscous bypass. However, increasing $S_{ws}$ has significant impact on water breakthrough and recovery, so we cannot maximize $\Lambda$---to avoid the need of a small-scale heterogeneity in the permeability field---and still match the production data of E2000.

In addition to the production data, we conclude that the key characteristics of an immiscible viscous fingering experiment are:
\begin{itemize}
    \item The dominant mode of the fingers before breakthrough;
    \item The presence or absence of a rarefaction zone trailing the fingers;
    \item The degree of tip splitting and tortuosity of the fingers;
    \item The presence or absence of a capillary induced speckled saturation field;
    \item The saturation distribution throughout the displacement, in particular the presence of bypassed regions at the late stage.
\end{itemize}

It should be noted that, if available, information about the onset of fingering, at the very early stage of the displacement, could be used to ground the choices for our flow functions by direct matching of the dominant mode observed with the $\nu_{max}$ from the LSA.\@ In practice, at the stage where a direct comparison with LSA results are possible, the fingers are still quite small, and imaging difficulties near the inlet may blur these early-stage fingers, as is the case in experiment E2000. In this section, we assume that such information is not available.

To characterize the instability of the displacement, we rely on an estimate of the dominant mode of the fingers in the nonlinear regime before breakthrough (item 1). After breakthrough, preferential pathways are established all the way to the outlet, and the balance of forces changes considerably, with capillary forces playing a more important role, and therefore is not an ideal period for identifying the instability of the displacement. We would like to capture the evolution of the dominant mode of the fingers throughout the displacement. While it is relatively straightforward to calculate the dominant wavenumber in vorticity fields from our simulations (see Section~S2 in \citealt{supmat}), the same does not apply to saturation field, i.e., x-ray scans of the experiment. One alternative is estimating $\hat{\nu}$ from the count of fingers in some portion of the porous medium, as we did in Section~\ref{sec:flowfunc_lsa}.

With all the above considerations in mind, we can now establish some guidelines for modelling the E2000 experiment, and other similar experiments:
\begin{itemize}
    \item Select flow functions that lead to a $\nu_{max}$ multiple times larger than the dominant wavenumber observed, $\hat{\nu}_{exp}$, in the experiment---matching $\nu_{max}$ to $\hat{\nu}_{exp}$ is most definitely conservative, and will likely lead to fingers that are too thick compared to the experiment; 
    \item The absence of a rarefaction zone, and the presence of tip splitting, are indications of small-scale (or finger-scale) heterogeneity in the permeability field;
    \item High variance in the small-scale heterogeneity can lead to a speckled saturation field, particularly if the displaced phase is the wetting phase and capillary heterogeneity effects are important; when the experiment does not show a speckled saturation field, the variance of the small-scale heterogeneity should be limited;
    \item The presence of bypassed oil regions (at scales larger than the finger scale) at the late stage of the displacement is an indication of larger-scale of heterogeneity in the permeability field; however, if the displaced phase is the wetting phase, capillary heterogeneity may lead to bypassed oil regions even with a relatively small contrast in the permeability field.
\end{itemize}

\section{Results}\label{sec:results}
In this section, we present a numerical model, created following the guidelines established in the previous section, which achieves a good match to the observed saturation fields and production data of the E2000 experiment. 

The flow functions chosen are the RP2 and Pc2 functions, which lead to a ${\nu_{max}{=}6\,\text{cm}^{-1}}$, around 10 times the dominant wavenumber estimated for E2000. As discussed above, we are constrained by the need to match the production data and our choice of RP2 is consistent with the production data of the experiment. For the capillary pressure, we choose the Pc2 function, which is a weakly oil-wet curve with relatively low ratio between dissipative and capillary-heterogeneity flow ($\left|J^{'}_c\right|/\left|J_c\right| \approx 1$, at the shock saturation). As we saw in Section~\ref{sec:capillary_heterogeneity}, a low $\left|J^{'}_c\right|/\left|J_c\right|$ ratio increases the impact of capillary heterogeneity compared to capillary diffusion, which in our case leads to increased bypass.

To match the observed saturation fields---both the absence of a rarefaction zone and the presence of bypassed oil---we use a \textit{multiscale} permeability field, where the high-frequency heterogeneity seeds the fingers, and the larger-scale heterogeneity promotes the formation of bypassed oil regions. The variance of the small-scale heterogeneity is limited to avoid a ``speckled'' saturation field. Because the slab used in E2000 is a relatively homogeneous porous medium, our goal is to reproduce the oil bypass with minimal contrast in the large-scale heterogeneity. We show that if capillary heterogeneity is present (and the displaced oil is the wetting phase), we can achieve this bypass with a relatively small contrast in the large-scale heterogeneity. 

The process of generating the multiscale permeability fields consists of first generating the larger-scale heterogeneity field via a Random Gaussian Simulation (RGS) with correlation length $\rho_{l1}$, and a log-normal distribution with median $k_{P50}$ and standard deviation $\sig_{\ln}$ (of $\ln(k)$). The permeability values are truncated at two standard deviations so that $k_{max}/k_{min}{=}\exp(4\sig_{\ln})$. Then, we generate a small-scale perturbation field using a second simulation with a smaller correlation length $\rho_{l2}$. This field has a normal distribution with mean zero and standard deviation of 0.5, truncated at two standard deviations, such that all values fall in the range $[-1, 1]$. The final permeability field is given by:
\begin{equation}\label{eq:multiscale_perm}
    k(x,y) = k_{l1}(x,y) \, m^{\delta_{l2}(x,y)},
\end{equation}
where $k_{l1}$ is the larger-scale field, $\delta_{l2}$ is the perturbation field, and $m$ controls the magnitude of the perturbation.

\begin{figure}[!htbp]
    \centering
    \includegraphics[width=0.5\textwidth]{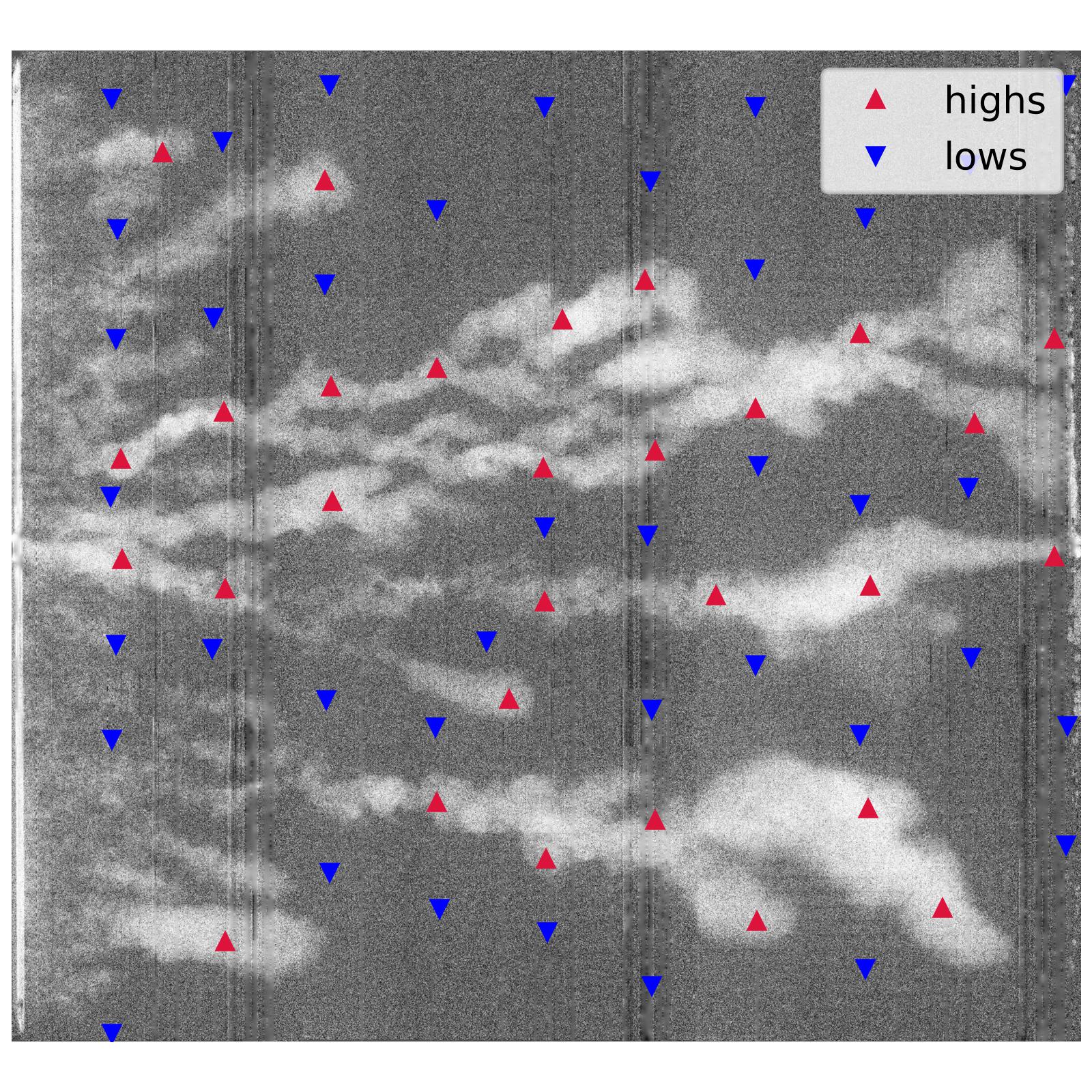}
    \caption{Conditioning data for the RGS of the larger-scale heterogeneity superimposed on the x-ray scan at 0.14 PVI.\@ The upward and downward triangles indicate the locations of high and low permeability, respectively. The high and low values are randomly selected between 1.5 and 2 standard deviations from the mean.}\label{fig:model1_conditioning}
\end{figure}

Instead of generating the larger-scale heterogeneity field with a completely random simulation, we use an ``image guide'' to condition the random simulation. We construct this guide by first identifying positions of high and low water saturation in the x-ray scan at 0.14 PVI, then randomly sampling high and low permeability values at these positions (between 1.5 and 2 standard deviations from the mean), and using these as conditioning data for the RGS of the large-scale heterogeneity. Figure~\ref{fig:model1_conditioning} shows the conditioning data extracted from the x-ray scan at 0.14 PVI. % chktex 13

\begin{figure}[!htbp]
    \centering
    \includegraphics[width=\textwidth]{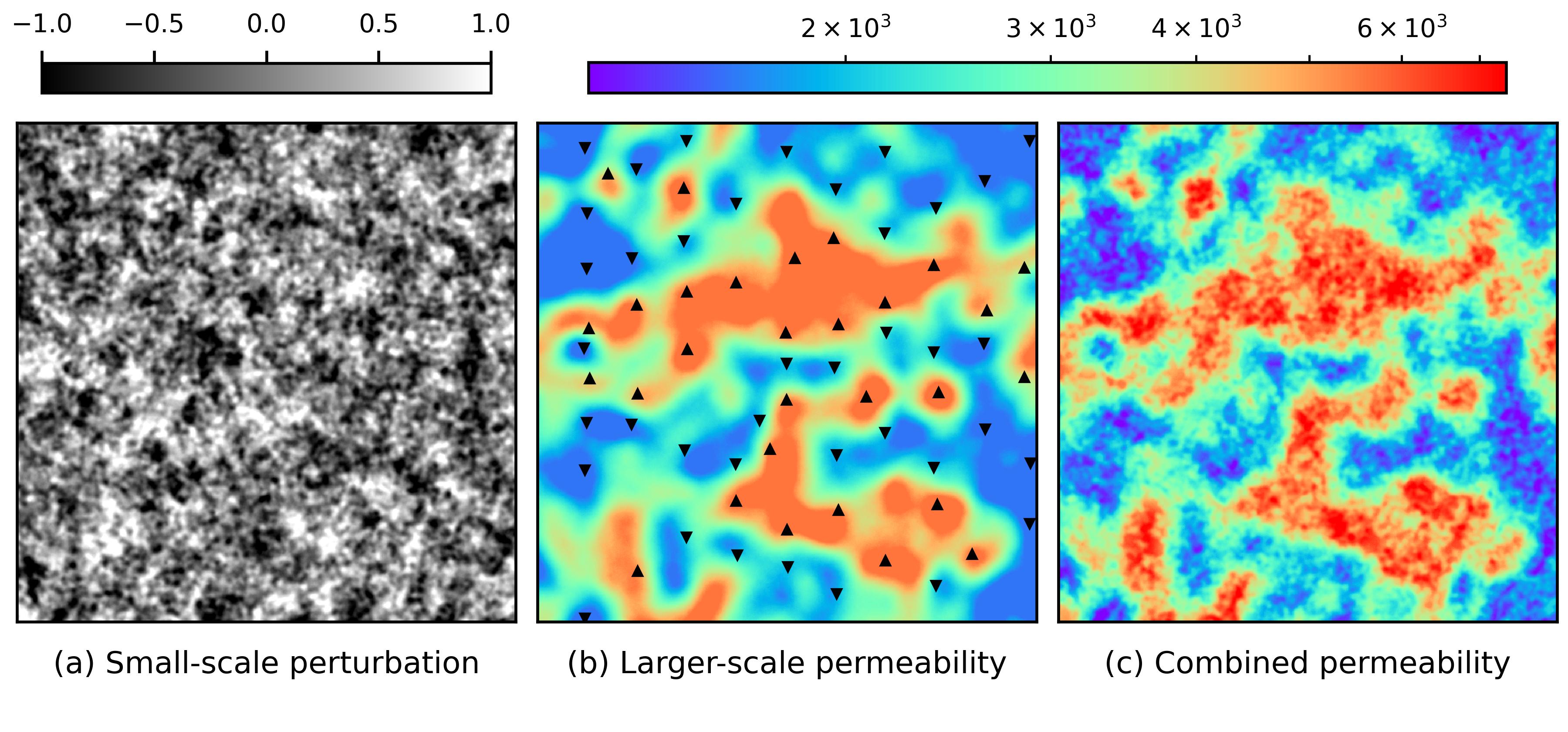}
    \caption{Perturbation and permeability (in mD) fields generated from the parameters in Table~\ref{tab:match_models}. In the larger-scale permeability map (b), the upward and downward triangles indicate the locations of high and low permeability, respectively, used to condition the RGS (see Figure~\ref{fig:model1_conditioning}).}\label{fig:model1_perm_maps}
\end{figure}

The matching process consists of a brute-force search over a range of values for the parameters of the multiscale permeability field ($\rho_{l1}$, $\sig_{\ln}$, $\rho_{l2}$ and $m$). Naturally, such a crude search cannot be expected to reproduce the exact saturation field, but our goal is to find a good match of the general characteristics of the saturation field, such as the presence of bypassed oil regions and the saturation distribution, for which many satisfactory matches are found. Here, we select one such model, summarized in Table~\ref{tab:match_models}. Figure~\ref{fig:model1_perm_maps} shows the small-scale perturbation, the larger-scale heterogeneity and the final multiscale permeability field for this model.

\begin{table}[!htbp]
    \centering
    \caption{Parameters of the selected multiscale permeability field.}\label{tab:match_models}
    \begin{tabular}{ccccccc}
        \toprule
        Image guide & $k_{P50}$ & $\sig_{\ln}$ & $k_{max}/k_{min}$ & $\rho_{l1}$ & $\rho_{l2}$ & $m$ \\ %[3pt]
        \midrule
        Figure~\ref{fig:model1_conditioning} & 3000 mD & 0.31 & 3.5 & 3 cm & 0.5 cm & 1.75 \\ % trial 13E
        \bottomrule
    \end{tabular}
\end{table}

\begin{figure}[!htbp]
    \centering
    \includegraphics[width=0.7\textwidth]{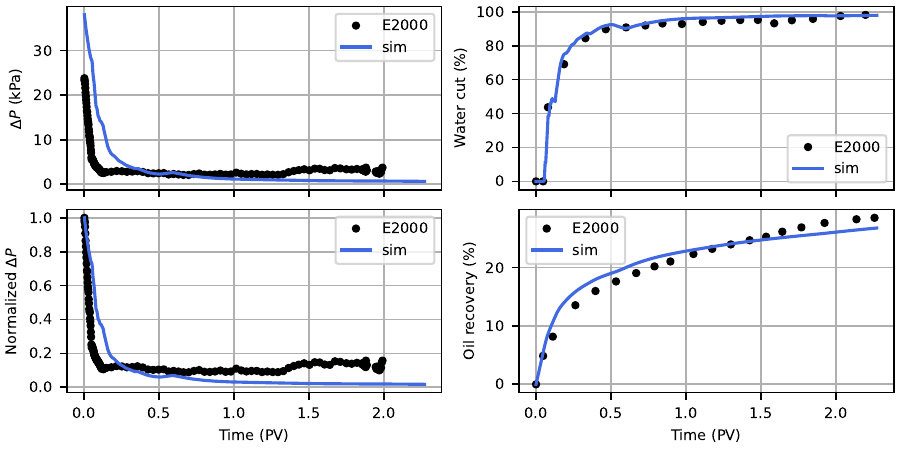}
    \caption{Experimental production data and simulation results. The top-left graph shows the pressure drop, the bottom-left graph shows the normalized pressure drop, the top-right shows the effluent water cut, and the bottom-right shows the oil recovery.}\label{fig:model1_prod_match}
\end{figure}

Figure~\ref{fig:model1_prod_match} shows the data and the simulation results for the normalized pressure drop, the effluent water cut and the oil recovery. The normalized pressure drop is defined as $\Delta P/\Delta P_{max}$, where $\Delta P$ is the difference between the inlet and outlet pressures, and $\Delta P_{max}$ is the maximum value of $\Delta P$. We can see that the multiscale model matches the production data well, capturing the moment of breakthrough and the general shape of the curves. 

\begin{figure}[!htbp]
    \centering
    \includegraphics[width=0.95\textwidth]{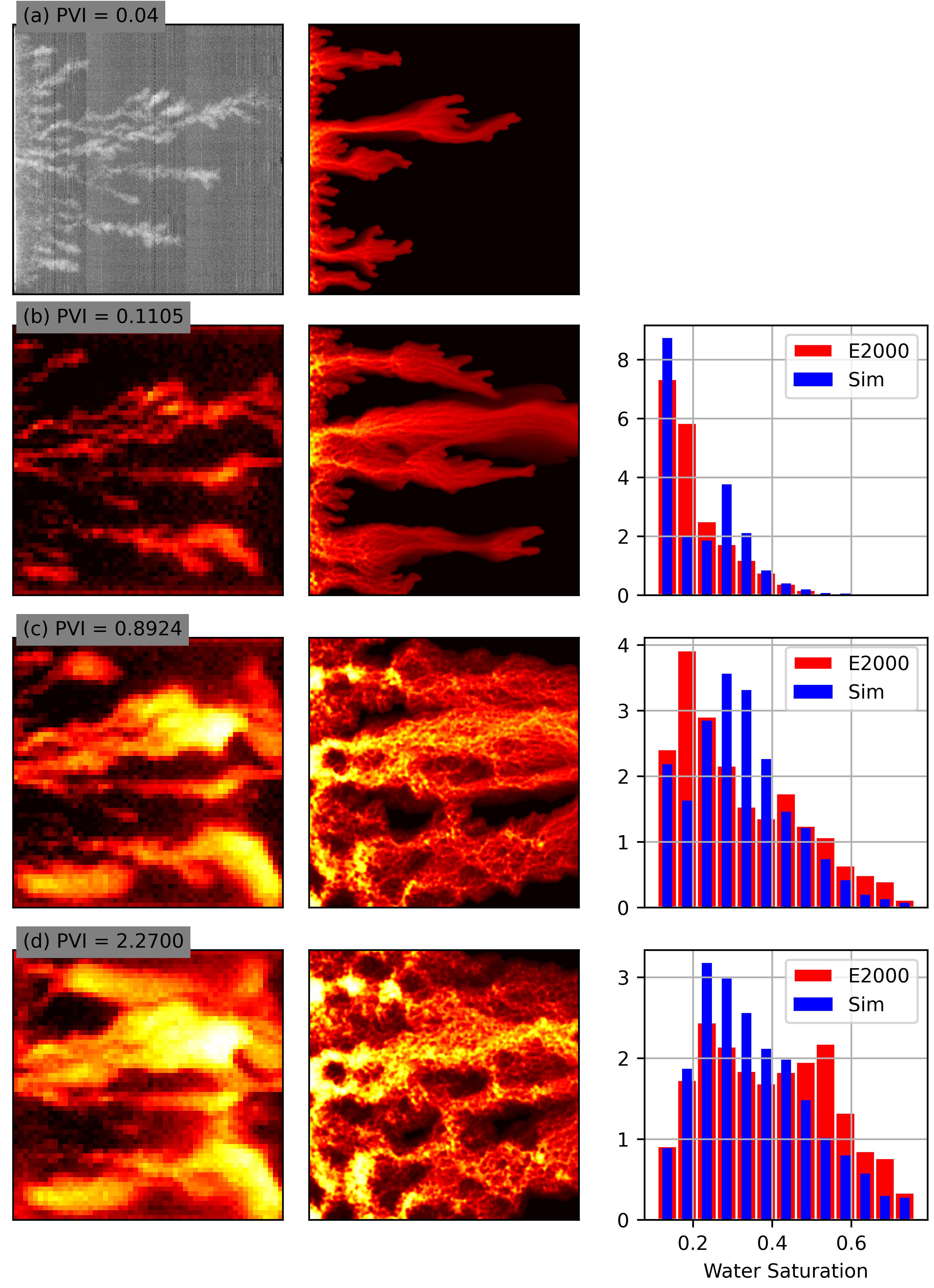}
    \caption{Saturation scans and simulation results. The left maps show the experimental data, the middle maps show the simulation results, and the right graphs show a comparison between the histograms for the experimental and simulated saturations. For the case at 0.04 PVI, only the ``fast'' scan was available, and so no histogram comparison is shown.}\label{fig:model1_sat_match}
\end{figure}

Figure~\ref{fig:model1_sat_match} shows the E2000 saturation data and the simulation results for multiscale model, at four different instants during the displacement: at 0.04 PVI (right before breakthrough), 0.11 PVI, 0.89 PVI and 2.3 PVI (end of the water-flood portion of the experiment). We note that two types of x-ray images were generated during the water-flood experiment: ``fast scans'' that took around 5 minutes (Figure~\ref{fig:xray_exp2000}), and ``saturation scans'' that took around 2 hours~\citep{skaugeT2014}, which we use to assess the match of the saturation distribution. Because of the long time required for the saturation scans, no such scan was taken before breakthrough, so for the pre-breakthrough time we provide only a visual comparison between the simulation and the fast scan. For the remaining instants, we show the saturation maps for the experiment and the simulation, as well as the experimental histograms of the water saturation.

\begin{figure}[!htbp]
    \centering
    \includegraphics[width=0.7\textwidth]{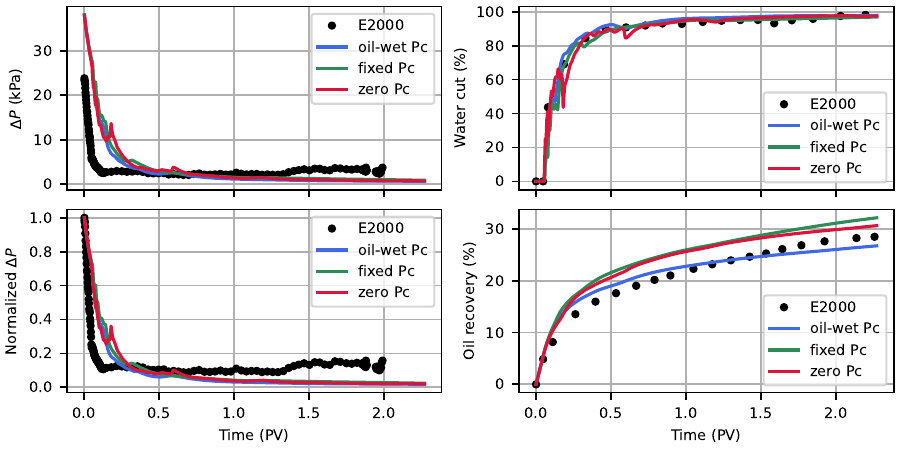}
    \caption{Same as Figure~\ref{fig:model1_prod_match} for the cases with: capillary heterogeneity (oil-wet Pc), homogeneous capillary pressure (fixed $P_c$), and no capillary pressure ($P_c{=}0$).}\label{fig:model1_prod_match_pcvar}
\end{figure}

\begin{figure}[!htbp]
    \centering
    \includegraphics[width=0.95\textwidth]{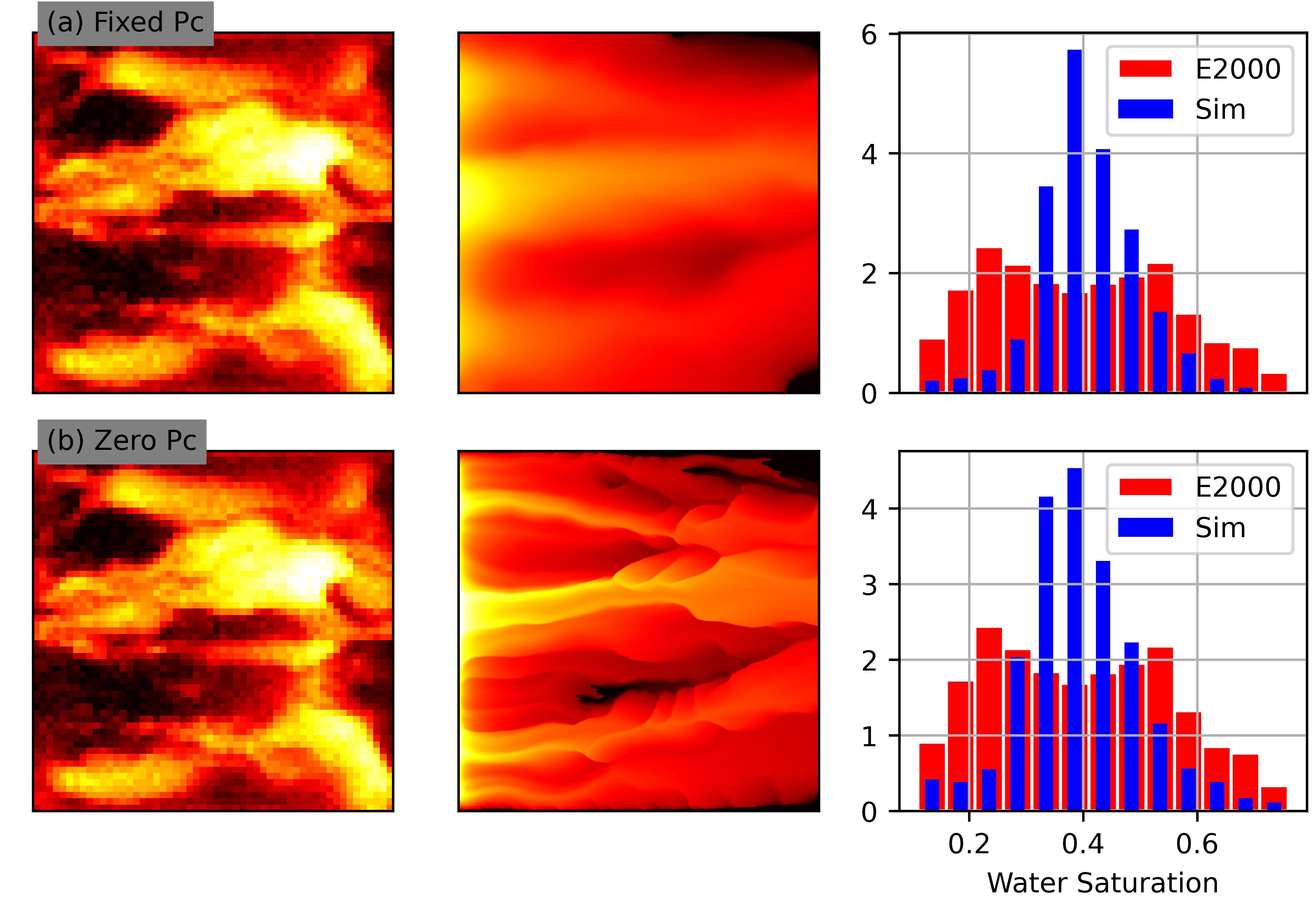}
    \caption{Saturation scans and simulation results, at 2.3 PVI, for the cases with fixed and zero $P_c$.}\label{fig:model1_sat_match_pcvar}
\end{figure}

We see a good qualitative match between the saturation maps our simulations, the multiscale model being able to capture the bypassed regions. We also observe a reasonably good match between the histograms. To show the impact of capillary heterogeneity in our model, we repeat the same simulation for two different cases: one where the $P_c$ is a function of saturation only (fixed at $k=3\,$D in Equation~\ref{eq:leverett}), and another without $P_c$ ($P_c{=}0$). Figures~\ref{fig:model1_prod_match_pcvar} and~\ref{fig:model1_sat_match_pcvar} show the match of the production and saturation data for these two cases. For both cases, the impact on the pressure drop observed is negligible, though we observe an increase in the oil recovery (of 20\% and 15\% for the fixed and zero $P_c$ cases, respectively). The difference in the saturation distribution is more apparent. Without capillary-heterogeneity effects, most of the medium is swept by the water phase at the end of the displacement (2.3 PVI), which explains the increase in recovery. The saturation histograms also show that the effect of capillary heterogeneity is to widen the saturation distribution: while preventing water from reaching the low-permeability regions (increasing the frequency of low saturation values), it concentrates the water in the high-permeability regions (increasing the frequency of high saturation values). Without this effect, the saturation distribution becomes narrower, as we see in Figure~\ref{fig:model1_sat_match_pcvar}.

The results shown in Figure~\ref{fig:model1_sat_match_pcvar} for the cases with fixed or zero $P_c$ highlights the main effect of capillary heterogeneity: it leads to a small-scale (but still macroscopic) trapping of the wetting phase (oil, in E2000) in the low-permeability regions. We argue that this is not only relevant for viscous fingering experiments at the laboratory scale, but also for field-scale displacements. Although at field scales the $P_c$ is usually considered negligible, its effect on recovery at the small-scale does transfer to the large-scale displacements, much like capillary trapping at the pore scale leads to residual saturations that directly influence Darcy-scale recovery.

Finally, we point out that the presence of bypassed oil in a relatively homogeneous rock is dependent on two factors: that ${\left| J_c' \right| / \left| J_c \right|}$ is small, and that oil is the wetting phase. This is the case for a weakly oil-wetting $P_c$ function, from which we modelled our $P_c$ function. If water was the wetting phase ($J_c{>}0$), capillary heterogeneity would promote the flow of water \textit{into} the low-permeability regions, while if ${\left| J_c' \right| / \left| J_c \right|}$ was large, then capillary-heterogeneity flow would be suppressed by the dissipative flow (provided $\nabla\tau/\tau$ was not too large). If, on the other hand, the injection rate was increased, then viscous forces would increase relative to capillary forces, and the impact of capillary heterogeneity would also be reduced, the limiting case being that of the zero $P_c$ case shown in Figure~\ref{fig:model1_sat_match_pcvar}. In these three cases, the presence of bypassed oil in a relatively homogeneous rock would be unlikely.

\FloatBarrier\section{Conclusions}\label{sec:conclusion}
In this study, we successfully developed a methodology, in the form of a set of guidelines, to model the nonlinear evolution of immiscible viscous fingering. We then applied this methodology to construct a numerical model that satisfactorily matches the main characteristics of the immiscible viscous fingering observed in the E2000 experiment. The match captures the fine scale of the fingers, the absence of a stable zone trailing the fingers, and the saturation distribution throughout the displacement, in particular the bypassed oil regions in the late displacement stages of the waterflood. The match to late time oil bypassing was evident both visually in the finger patterns and also in reproducing the experimental distributions (histograms) of the remaining oil.

Our simulations showed that capturing these characteristics requires:
\begin{enumerate}
    \item~The most unstable wavenumber ($\nu_{max}$, at the onset of instability) to be much larger than the dominant wavenumber in the nonlinear regime, such that, even after accounting for shielding and merging effects, fingers are similar to experimentally observed ones; 
    \item~Small-scale heterogeneity in the permeability field to disrupt the rarefaction zone trailing the fingers; 
    \item~Larger-scale heterogeneity in the permeability field to promote the formation of bypassed oil regions at the late stage of the displacement.
\end{enumerate}

We showed that with a weakly oil-wet capillary pressure curve (oil being the displaced phase), where relatively small ratio of $\left| J_c' \right| / \left| J_c \right|$ means capillary heterogeneity effects are not suppressed by capillary diffusion, the formation of bypassed oil regions happens even with a relatively small contrast in the permeability field, which can have a significant impact on the displacement efficiency, even for relatively homogeneous porous media. 

In our study of the evolution of fingers in a homogeneous medium, we showed that if the initial wavelike perturbation maintains its symmetry (i.e., fingers have the same length as their neighbours), then nonlinear effects---mainly due to an increasingly corrugated displacement front, which promotes capillary-dissipative flow---limit finger growth to a maximum proportional to their wavelength (in an apparent power-law relation, for the case studied; \citealt{supmat}). We interpret this as an additional mechanism, besides the shielding effects discussed by \citet{riaz2006b}, that leads to a decrease in the dominant wavenumber of the fingers. Simply put, thicker fingers are allowed to grow longer.

%%%% Acknowledgements and statements %%%%
\backmatter%

\bmhead{Supplementary information} The following supplementary material is available for this article: \citet{supmat}---sections include: (S1)~Linear stability analysis; (S2)~Growth rate and finger size calculation; (S3)~Nonlinear evolution of immiscible fingers.

\bmhead{Acknowledgements} The authors acknowledge the PhD scholarship for Paulo L. K. Caetano Chang provided by Petrobras. Kundan Kumar acknowledges funding from the Centre of Sustainable Subsurface Resources (CSSR), grant nr.\ 331841, supported by the Research Council of Norway, research partners NORCE Norwegian Research Centre and the University of Bergen, and user partners Equinor ASA, Harbour Energy Norge AS, Sumitomo Corporation, Earth Science Analytics, GCE Ocean Technology, and SLB Scandinavia. Arne Skauge acknowledges Energi Simulation for support of the chair in Low Net-Carbon EOR and Energy Transition at Heriot-Watt University, Edinburgh, UK.\@ Tormod Skauge is thanked for his helpful input.

\section*{Declarations}
\bmhead{Funding} Paulo Lee Kung Caetano Chang received funding from Petrobras. Kundan Kumar received funding from Centre of Sustainable Subsurface Resources, Grant ID 331841.

\bmhead{Competing interests} The authors have no competing interests to declare that are relevant to the content of this article.

\bmhead{Ethics approval} Not applicable.

\bmhead{Consent for publication} Not applicable.

\bmhead{Data availability} The data that support the findings of this study are available upon reasonable request.  % chktex 12

\bmhead{Materials availability} Not applicable.

\bmhead{Code availability} Not applicable.

% Using the Contributor Role Taxonomy (CRediT)
\bmhead{Author contribution} \textbf{P. L. K. Caetano Chang}: conceptualization; methodology; writing---original draft; writing---review \& editing. \textbf{K. Kumar}: conceptualization; writing---review \& editing; supervision. \textbf{A. Skauge}: conceptualization; methodology; writing---review \& editing; supervision. \textbf{K. S. Sorbie}: conceptualization; writing---review \& editing; supervision.

%%%% References %%%%
\bibliography{references.bib}% common bib file

\end{document}

\maketitle

\section{Linear stability analysis (LSA)}\label{sec:linearstab}
The main goal of LSA is to describe the onset (or suppression) of instability in a two-phase displacement in a porous medium. In this technique modal analysis is applied around the linearized governing equations of the flow, to determine the growth rate of small perturbations as a function of their wavenumber. The result is called the dispersion relation, and it can be used to determine the stability of the system.

The first step is to define the base-state solution, which is usually a steady-state solution of the unperturbed displacement. Then, the governing equations are linearized for small variations around the base-state solution to obtain the linearized perturbed equations. The modal analysis is then performed by decomposing the dependent variables into a base-state and small wavelike perturbations---that can grow or decay---and, after substituting these into the linearized equations, we arrive at the eigenvalue problem in terms of the disturbances.

In this section, we describe the main equations of the LSA for the incompressible, two-phase immiscible displacement in a homogeneous porous medium, as developed by \citet{riaz2004}. We consider only the case of horizontal flow (i.e.\ no gravity effects). The governing equations are well known and given by:
\begin{align}
    &\phi \frac{\partial S_\al}{\partial t} + \nabla \cdot \bol{u}_\al = 0, \label{eq:matbal} \\
    &\bol{u}_\al = -k \frac{k_{r\al}}{\mu_\al} \nabla P_\al, \label{eq:darcy} \\
    &\nabla \cdot \left( \bol{u}_w + \bol{u}_o \right) = 0, \label{eq:continuity} \\
    &P_c = P_o - P_w, \label{eq:cap_pressure}
\end{align}
where $\bol{u}_\al$ is the Darcy velocity of phase $\al(=w,o)$, $\phi$ is the porosity, $k$ is the absolute permeability, $k_{r\al}$ is the relative permeability of phase $\al$, $\mu_\al$ is the viscosity of phase $\al$, $P_\al$ is the pressure of phase $\al$, and $P_c$ is the capillary pressure.

The equations can be nondimensionalized by the following transformations:
\begin{equation}\label{eq:normalization}
    \begin{aligned}
        \bol{x}^*=&\frac{\bol{x}}{L}, &\bol{u}^*=&\frac{\bol{u}}{U}, &t^*=&\frac{U}{\phi L}t,\\
        \la_\al^*=&k_{r\al}, &P_\al^*=&\frac{k}{U\mu_\al L}P_\al, &P_c^*=&\frac{k}{U\mu_w L}P_c,\\ 
    \end{aligned}
\end{equation}
where $U$ is the injection specific discharge (or Darcy velocity).

Equations~\ref{eq:matbal}--\ref{eq:cap_pressure} can be rewritten as:
\begin{align}
    &\frac{\partial S_\al}{\partial t^*} + \nabla_{*} \cdot \bol{u}_\al^* = 0, \label{eq:matbal*} \\
    &\bol{u}_\al^* = \la_\al^* \nabla_{*} P_\al^*, \label{eq:darcy*} \\
    &\nabla_{*} \cdot \left( \bol{u}_w^* + \bol{u}_o^* \right) = 0, \label{eq:continuity*} \\
    &P_c^* = M P_o^* - P_w^*, \label{eq:cap_pressure*}
\end{align}
where $M = \mu_o/\mu_w$ is the viscosity ratio and $\nabla_* = \partial/\partial x_i^*$.

For convenience, we omit the asterisks from here on, with the understanding that all variables are now dimensionless.

The one-dimensional, unperturbed reference state (i.e.\ the base state), around which the flow equations are linearized, is derived from the moving frame formulation of the Buckley-Leverett equation, given by:
\begin{equation}\label{eq:basestate_eq}
    \frac{\partial S_w}{\partial t} + \left(\frac{\mathrm{d}F_w}{\mathrm{d}S_w} - V_s\right) \frac{\partial S_w}{\partial \xi} + \frac{\partial}{\partial \xi} \left( \frac{\la_w \la_o}{\la_t} \frac{\mathrm{d}P_c}{\mathrm{d}S_w} \frac{\partial S_w}{\partial \xi} \right) = 0,
\end{equation}
where $\la_t=M\la_w+\la_o$ is the total mobility, $\xi=x - V_s t$ is the moving frame coordinate that moves with the shock front velocity $V_s=dF_w/dS_w(S_{ws})$, $S_{ws}$ being the shock saturation, and $F_w$ is the fractional flow function of the water phase,
\begin{equation}\label{eq:fw}
    F_w = \frac{M \la_w}{\la_t} = \frac{M k_{rw}}{M k_{rw} + k_{ro}}.
\end{equation}

The base-state solution is the steady-state solution of Equation~\ref{eq:basestate_eq}, given by:
\begin{equation}\label{eq:basestate_sol}
    D_c \frac{\mathrm{d}S_w}{\mathrm{d}\xi} = V_s (S_w - S_{w0}) + F_w(S_{w0}) - F_w(S_w),
\end{equation}
where $S_{w0}$ is the initial water saturation and $D_c$ is the dispersion coefficient given by:
\begin{equation}\label{eq:Dc}
    D_c = \frac{\la_w \la_o}{\la_t} \frac{\mathrm{d}P_c}{\mathrm{d}S_w}.
\end{equation}

The base-state pressure is given by:
\begin{equation}\label{eq:basestate_pressure}
    \la_o \frac{\mathrm{d}P_o}{\mathrm{d}S_w} = V_s \left(S_w - S_{w0}\right) + F_w(S_{w0}) - 1.
\end{equation}

The perturbed equations in terms of the water saturation and oil pressure are:
\begin{align}
    \frac{\partial S_w}{\partial t} - V_s \frac{\partial S_w}{\partial \xi} + \nabla \cdot \left( \la_o \nabla P_o \right) &= 0, \label{eq:perturbed_matbal} \\
    \nabla \cdot \left( \la_t \nabla P_o - \la_w P_c'\nabla S_w \right) &= 0. \label{eq:perturbed_continuity}
\end{align}

To proceed with the LSA, we decompose the perturbed variables $S_w$ and $P_o$ into a base state and a small wavelike perturbation as follows:
\begin{align}
    S_w(\xi,y,t) &= \bar{S}(\xi) + \hat{s}(\xi) \cos(2\pi\nu y) e^{\sig t}, \label{eq:decomp_S} \\
    P_o(\xi,y,t) &= \bar{P}(\xi) + \hat{p}(\xi) \cos(2\pi\nu y) e^{\sig t}, \label{eq:decomp_P}
\end{align}
where $\bar{S}$ and $\bar{P}$ are the solutions of the base-state Equations~\ref{eq:basestate_sol} and~\ref{eq:basestate_pressure}, respectively, $\nu$ is the wavenumber of the perturbation, and $\sig$ is its growth rate. We assume the disturbances decay at infinity, i.e. $\hat{s},\hat{p} \to 0$ for $\xi \to \pm\infty$.

Substituting Equations~\ref{eq:decomp_S} and~\ref{eq:decomp_P} into Equations~\ref{eq:perturbed_matbal} and~\ref{eq:perturbed_continuity} and linearizing the flow functions (i.e.\ $\la_\al$ and $P_c$) around $\bar{S}$, we arrive, after considerable algebra, at the eigenvalue problem:
\begin{equation}\label{eq:eigenvalue_problem}
    \begin{bmatrix}
        A\ & B \\
        C\ & D-\sig
    \end{bmatrix} 
    \begin{Bmatrix}
        \hat{p} \\ 
        \hat{s}
    \end{Bmatrix}
        = 0,
\end{equation}
where $A$, $B$, $C$, and $D$ are linear operators of the type $L = L_2 \tfrac{d^2}{d\xi^2} + L_1 \tfrac{d}{d\xi} + L_0$. The coefficients $A_0,A_1,\ldots$ are given by:
\begin{equation}\label{eq:coefficients}
    \begin{aligned}
        A_2 =&\ \la_t, \\
        A_1 =&\ \la_t' \frac{\mathrm{d}\bar{S}}{\mathrm{d}\xi}, \\
        A_0 =&\ -4\pi^2\nu^2 \la_t,\\
        B_2 =&\ -\la_w P_c', \\
        B_1 =&\ \la_t'\frac{\mathrm{d}\bar{P}}{\mathrm{d}\xi} - 2\left( \la_w'P_c' + \la_w P_c'' \right)\frac{\mathrm{d}\bar{S}}{\mathrm{d}\xi}, \\
        B_0 =&\ \la_t'\frac{\mathrm{d}^2\bar{P}}{\mathrm{d}\xi^2} + \la_t''\frac{\mathrm{d}\bar{S}}{\mathrm{d}\xi}\frac{\mathrm{d}\bar{P}}{\mathrm{d}\xi} - \left( \la_w''P_c' + 2\la_w'P_c'' + \la_w P_c^{\prime\prime\prime} \right){ \left( \frac{\mathrm{d}\bar{S}}{\mathrm{d}\xi} \right)}^2 \\
             &- \left( \la_w'P_c' + \la_w P_c'' \right) \frac{\mathrm{d}^2\bar{S}}{\mathrm{d}\xi^2} + 4\pi^2\nu^2\la_w P_c', \\
        C_2 =&\ -\la_o, \\
        C_1 =&\ -\la_o' \frac{\mathrm{d}\bar{S}}{\mathrm{d}\xi}, \\
        C_0 =&\ 4\pi^2\nu^2 \la_o,\\
        D_2 =&\ 0, \\
        D_1 =&\ v_s - \la_o'\frac{\mathrm{d}\bar{P}}{\mathrm{d}\xi}, \\
        D_0 =&\ -\la_o''\frac{\mathrm{d}\bar{S}}{\mathrm{d}\xi}\frac{\mathrm{d}\bar{P}}{\mathrm{d}\xi} - \la_o'\frac{\mathrm{d}^2\bar{P}}{\mathrm{d}\xi^2} 
    \end{aligned}
\end{equation}

We can solve the system in Equation~\ref{eq:eigenvalue_problem} by discretizing the linear operators using a finite difference:
\begin{equation}\label{eq:eigenvalue_discrete}
    \begin{bmatrix}
        \mathsfbi{A}\ & \mathsfbi{B} \\
        \mathsfbi{C}\ & \mathsfbi{D}-\sig\mathsfbi{I}
    \end{bmatrix} 
    \begin{Bmatrix}
        \hat{\bol{p}} \\ 
        \hat{\bol{s}}
    \end{Bmatrix}
        = \bol{0},
\end{equation}
where $\mathsfbi{A}$, $\mathsfbi{B}$, $\mathsfbi{C}$, and $\mathsfbi{D}$ are square matrices of size $N\times N$, with $N$ being the number of grid points in the $\xi$ direction, and $\hat{\bol{p}}$ and $\hat{\bol{s}}$ are column vectors of size $N$ with the discrete solution of $\hat{p}$ and $\hat{s}$. Assuming that $\mathsfbi{A}$ is invertible, we can solve for $\hat{\bol{p}}$ to obtain:
\begin{equation}\label{eq:sol_p}
    \hat{\bol{p}} = -\mathsfbi{A}^{-1}\mathsfbi{B}\hat{\bol{s}}.
\end{equation}

Substituting Equation~\ref{eq:sol_p} into the second row of Equation~\ref{eq:eigenvalue_discrete} we arrive at the standard eigenvalue problem:
\begin{equation}\label{eq:eigenvalue_standard}
    \mathsfbi{M}\hat{\bol{s}} = \sig \hat{\bol{s}},\quad \mathsfbi{M} = \mathsfbi{D} - \mathsfbi{C}\mathsfbi{A}^{-1}\mathsfbi{B}.
\end{equation}

The dispersion relation is given by the largest positive eigenvalue $\sig$ of $\mathsfbi{M}$ for every possible value for wavenumber $\nu$, and for the unbounded domain, that is any real value. For the case where flow is bounded laterally by two impermeable walls at $y=\pm L_y/2$, the no-flux boundary conditions must be satisfied, which translates to: 
\begin{equation}\label{eq:lateralwall_bc}
    \left. \frac{\partial S_w}{\partial y} \right|_{y=\pm \tfrac{L_y}{2}} = \left. \frac{\partial P_o}{\partial y}  \right|_{y=\pm \tfrac{L_y}{2}} = 0.
\end{equation}

Satisfying the no-flux boundary conditions limits the possible values for $\nu$ to:
\begin{equation}\label{eq:nu_restricted}
    \nu = \frac{n}{L_y},\quad n=1,2,3,\dots\ .
\end{equation}

For numerical simulations of fingers seeded by wavelike perturbations, the perturbation wavenumbers must be restricted to those given by Equation~\ref{eq:nu_restricted} so that the simulation conditions match those of the LSA.\@

\section{Growth rate and finger size calculation}\label{sec:fingercalc}
To calculate the growth rate of the fingers, we turn to the vorticity field, ${\bol{\omega}{=}\nabla {\times} \bol{u}_t}$, which for the two-dimensional homogeneous case (no gravity) can be written as:
\begin{equation}\label{eq:vorticity}
    \bol{\omega} = \frac{1}{\la_t} \frac{\mathrm{d}\la_t}{\mathrm{d}S_w} \nabla S_w \times \bol{u}_t = -\frac{\mathrm{d}\la_t}{\mathrm{d}S_w} \nabla S_w \times \nabla P_o.
\end{equation}

Substituting the perturbed variables $S_w$ and $P_o$ (Equations~\ref{eq:decomp_S} and~\ref{eq:decomp_P}) into Equation~\ref{eq:vorticity} and linearizing the flow functions around the base state, we can write the vorticity field as:
\begin{equation}\label{eq:vorticity_perturbed}
    \bol{\omega} = 2\pi\nu \frac{\mathrm{d}\la_t}{\mathrm{d}S_w} \left( \frac{\mathrm{d}\bar{S}}{\mathrm{d}\xi}\hat{p} - \frac{\mathrm{d}\bar{P}}{\mathrm{d}\xi}\hat{s} \right) \sin(2\pi\nu y) e^{\sig t} \bol{\hat{z}} = \hat{\omega}(\xi,y) e^{\sig t} \bol{\hat{z}},
\end{equation}
where $\bol{\hat{z}}$ is the unit vector in the $z$ direction normal to the plane of flow. We can see that the vorticity field has the same growth rate as the fingers at the onset of the instability. 

With the norm of the vorticity field defined as \citep{riaz2003}:
\begin{equation}\label{eq:vorticity_norm}
    \lVert\bol{\omega}\rVert (t) = {\left( \int_{-\tfrac{L_y}{2}}^{\tfrac{L_y}{2}} \int_{-\infty}^{\infty} {\left( \hat{\omega}(\xi,y) e^{\sig t} \right)}^2 d\xi dy \right)}^{\tfrac{1}{2}} = e^{\sig t} \lVert\hat{\omega}(\xi,y)\rVert,
\end{equation}
the growth rate of the fingers can then be calculated by:
\begin{equation}\label{eq:growth_rate_0}
    \sig = \frac{\mathrm{d}}{\mathrm{d}t} \ln \lVert\bol{\omega}\rVert,
\end{equation}

Although our simulations are formulated in the fixed frame $(x,y)$ rather than the moving frame $(\xi,y)$, $\hat{\omega}$ is nonzero only in the vicinity of the shock front---where the eigenfunctions $\hat{p}$ and $\hat{s}$ are nonzero---so $\lVert\bol{\omega}\rVert$ can be computed as:
\begin{equation}\label{eq:avg_vorticity}
    \lVert\bol{\omega}\rVert = \sqrt{\sum_{i=1}^{N_x} \sum_{j=1}^{N_y} {\omega_z(x_i,y_j,t)}^2},
\end{equation}
where $w_z$ is the $z$-component of the vorticity field given by Equation~\ref{eq:vorticity}, 
and the growth rate of the perturbations can be computed as:
\begin{equation}\label{eq:growth_rate}
    \sig \approx \frac{1}{\Delta t} \ln \left( \frac{\lVert\bol{\omega}\rVert(t)}{\lVert\bol{\omega}\rVert(t-\Delta t)} \right),
\end{equation}

As shown by \citet{riaz2006b}, the dominant mode (or wavenumber), $\hat{\nu}$, of the fingers can be estimated by:
\begin{equation}\label{eq:dominant_mode}
    \hat{\nu} = \frac{ \int_c^\nu \nu E(\nu, t) d\nu}{ \int_c^\nu E(\nu, t) d\nu}
\end{equation}
where $E(\nu, t)$ is the energy spectrum of the disturbed flow, calculated via the Fourier transform of the longitudinally averaged vorticity field given by:
\begin{equation}\label{eq:vorticity_fourier}
    E(\nu, t) = {\left[\int_{-\tfrac{L_y}{2}}^{\tfrac{L_y}{2}} \left( \int_0^{L_x} \omega_z(x,y,t) dx \right) e^{-\mathrm{i} 2\pi \nu y} dy \right]}^2
\end{equation}

\begin{figure}[!ht]
    \centering
    \includegraphics[width=0.5\textwidth]{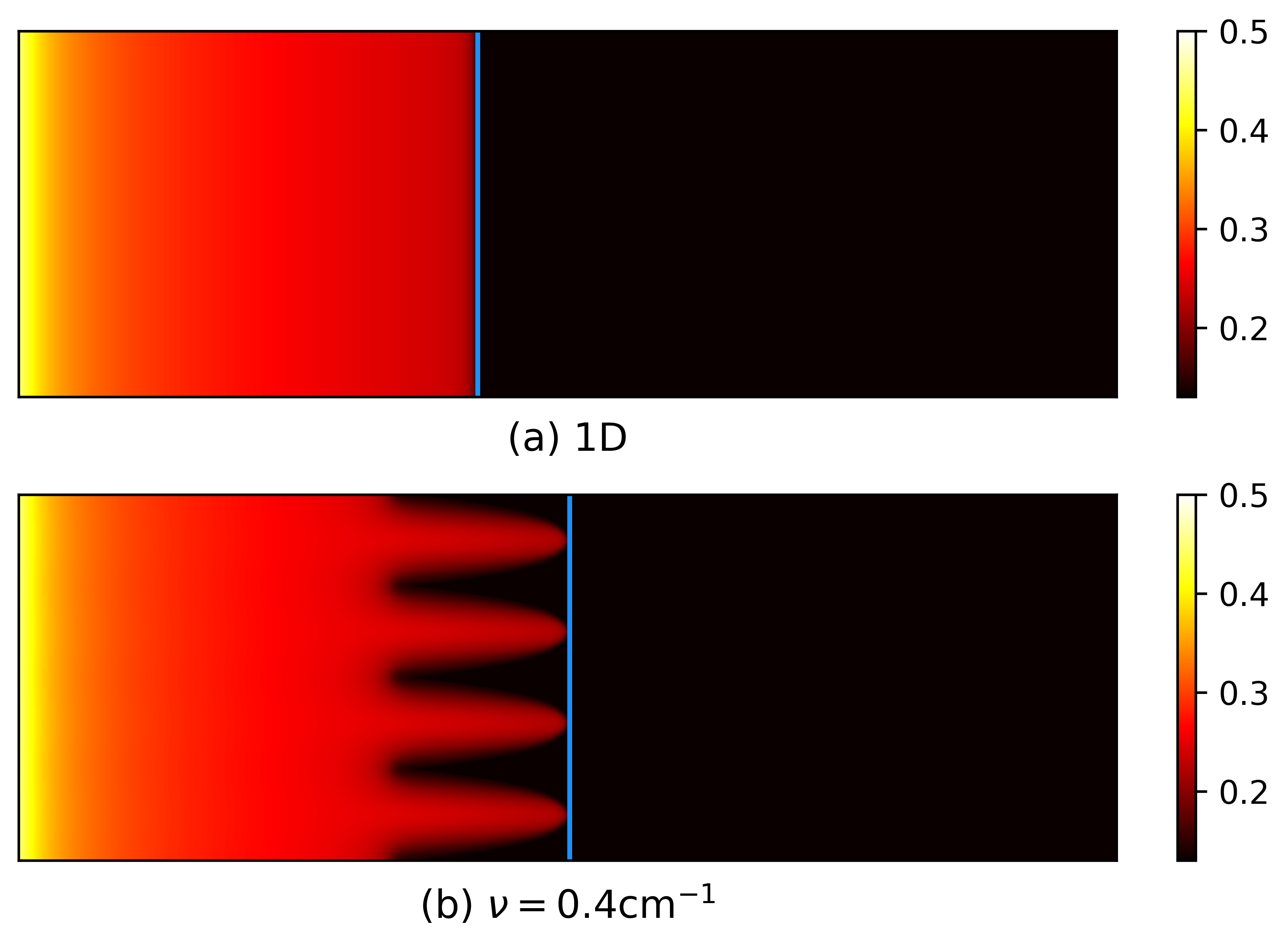}
    \caption{Saturation maps for the (unperturbed) one-dimensional case and the case with wavelike perturbations ($\nu=0.4$ cm$^{-1}$). The blue lines indicate the front most position of the displacements. The distance between the two fronts, at any given instant, gives the finger size.}\label{fig:cossw_size_calc}
\end{figure}

Another useful variable for tracking the evolution of the fingers is the \textit{finger~size}\footnote{An obvious way to calculate the growth rate of the fingers is to simply apply Equation~\ref{eq:growth_rate} to finger size, instead of the vorticity norm. In practice, though, at the early stages, finger size evolution is not an accurate measure, even for very fine grids, because the fingers are still very small. So using the vorticity norm is preferred.}, which we define as the distance between the front of the fingers and the front of the unperturbed case (i.e., the solution to the one-dimensional, Buckley-Leverett equation), as shown in Figure~\ref{fig:cossw_size_calc}.

\FloatBarrier\section{Nonlinear evolution of immiscible fingers}\label{sec:fingerevol}
In this section, we analyse the nonlinear evolution of immiscible fingers generated by wavelike perturbations in a homogeneous porous medium. The model has dimensions ${L_x=30\,\text{cm}}$ and ${L_y=10\,\text{cm}}$, with cell dimensions of ${l_x=l_y=0.02\,\text{cm}}$. For the five rows of cells nearest the inlet (at $x=0$), the saturation is set to $S_w{=}S_{wf}{-}0.005 \cos(2\pi \nu y)$, where $S_{wf}$ is the water saturation at the shock front and $\nu$ is the wavenumber of the perturbation ($\nu=n/L_y$, $n=1,2,\ldots$). For the remaining cells, the initial saturation is set to $S_w = S_{wr}$ (the residual water saturation). The RP and capillary pressure functions used in this simulation are RP1 and Pc1, respectively.

\begin{figure}[!ht]
    \centering
    \subfloat[Growth rate\label{fig:cossw_sigma_krv6a_2}]{
        \includegraphics[width=0.48\textwidth]{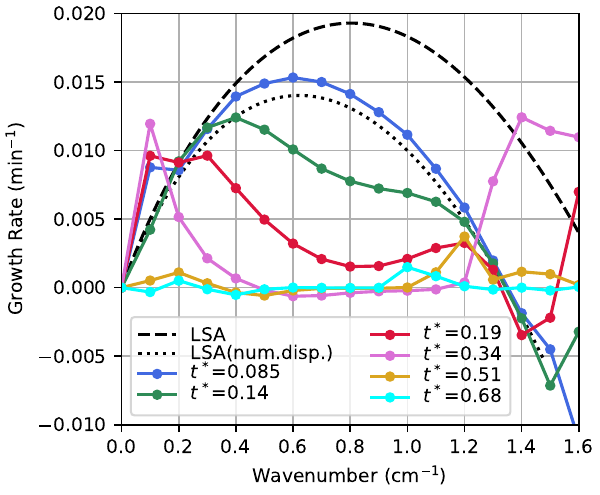}
    }\hfill
    \subfloat[Finger size\label{fig:cossw_size_krv6a_2}]{
        \includegraphics[width=0.48\textwidth]{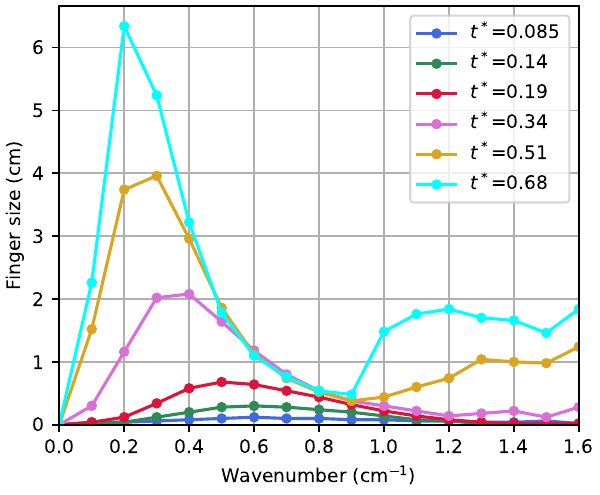}
    }
    \caption{Growth rate and size of the fingers at different times, for different wavenumbers, for the homogeneous model. The time for breakthrough $(t_{bt})$ is 1770 min and corresponds to 0.156 PVI.\@}\label{fig:cossw_size_sigma_krv6a_2}
\end{figure}

Figure~\ref{fig:cossw_sigma_krv6a_2} compares the growth rate of the fingers in the numerical simulations with the growth rate given by the LSA\@ The dashed line plots the dispersion relation given by the LSA, and the dotted line is the dispersion relation when we account for the extra dissipative effects of numerical dispersion on the base-state solution (Equation~\ref{eq:basestate_sol}), which is done by adding a numerical diffusion term to the dispersive coefficient $D_c$ (Equation~\ref{eq:Dc}) to obtain:
\begin{equation}\label{eq:Dcnum}
    D_c^{num} = D_c + D_{num} = \frac{\la_w \la_o}{\la_t} \frac{\mathrm{d}P_c}{\mathrm{d}S_w} - \frac{1}{2} \frac{\mathrm{d}F_w}{\mathrm{d}S_w} \left( \Delta x + \frac{\mathrm{d}F_w}{\mathrm{d}S_w} \Delta t \right),
\end{equation}
where $\Delta x$ is the cell size and $\Delta t$ is the maximum time step used in the numerical simulations (in non-dimensional units). Equation~\ref{eq:Dcnum} assumes the backward difference, implicit in time scheme \citep{lantz1971}. 

We see that the corrected dispersion relation\footnote{Note that only the base-state solution is corrected, while the coefficients of the linear operators in the eigenvalue problem (Equation~\ref{eq:coefficients}) remain unchanged.}\footnote{Although we do not show it here, we observe that the impact of numerical dispersion diminishes in the non-linear regime, and so its impact on finger evolution is less significant.} (dotted line) matches the early-time ($t^*=0.085$) results of the numerical simulations. As the simulation progresses, nonlinear effects become more important and the growth rate of the fingers decreases, eventually reaching zero---i.e., the fingers reach a steady state.

Figure~\ref{fig:cossw_size_krv6a_2} shows how the fingers grow in length over time for each initial wavenumber. At the onset, when the perturbations are small and their growth rates can be well described by the LSA, the size of the fingers is proportional to the growth rate of the initial perturbation. As the simulation progresses, nonlinear effects become more important and the fingers generated from smaller wavenumbers (i.e., larger wavelengths) keep growing, while those generated from larger wavenumbers stagnate at sizes inversely proportional to their wavenumbers.

This is consistent with the weakly nonlinear analysis by \citet{chikhliwala1988b}, which found that perturbations near the critical wavenumber ($\nu_{cut}$, from linear stability) are supercritical, stabilizing to an equilibrium amplitude (proportional to the mobility ratio at the displacement front). In our simulations, we find that this supercritical stability is not only true in the vicinity of $\nu_{cut}$, but, possibly, for all wavenumbers. Later in this section, we discuss finger stagnation in more detail.

\begin{figure}[!ht]
    \centering
    \subfloat[$\nu{=}0.1\,\text{cm}^{-1}$\label{fig:cossw_vort_maps_a}]{
        \includegraphics[width=0.48\textwidth]{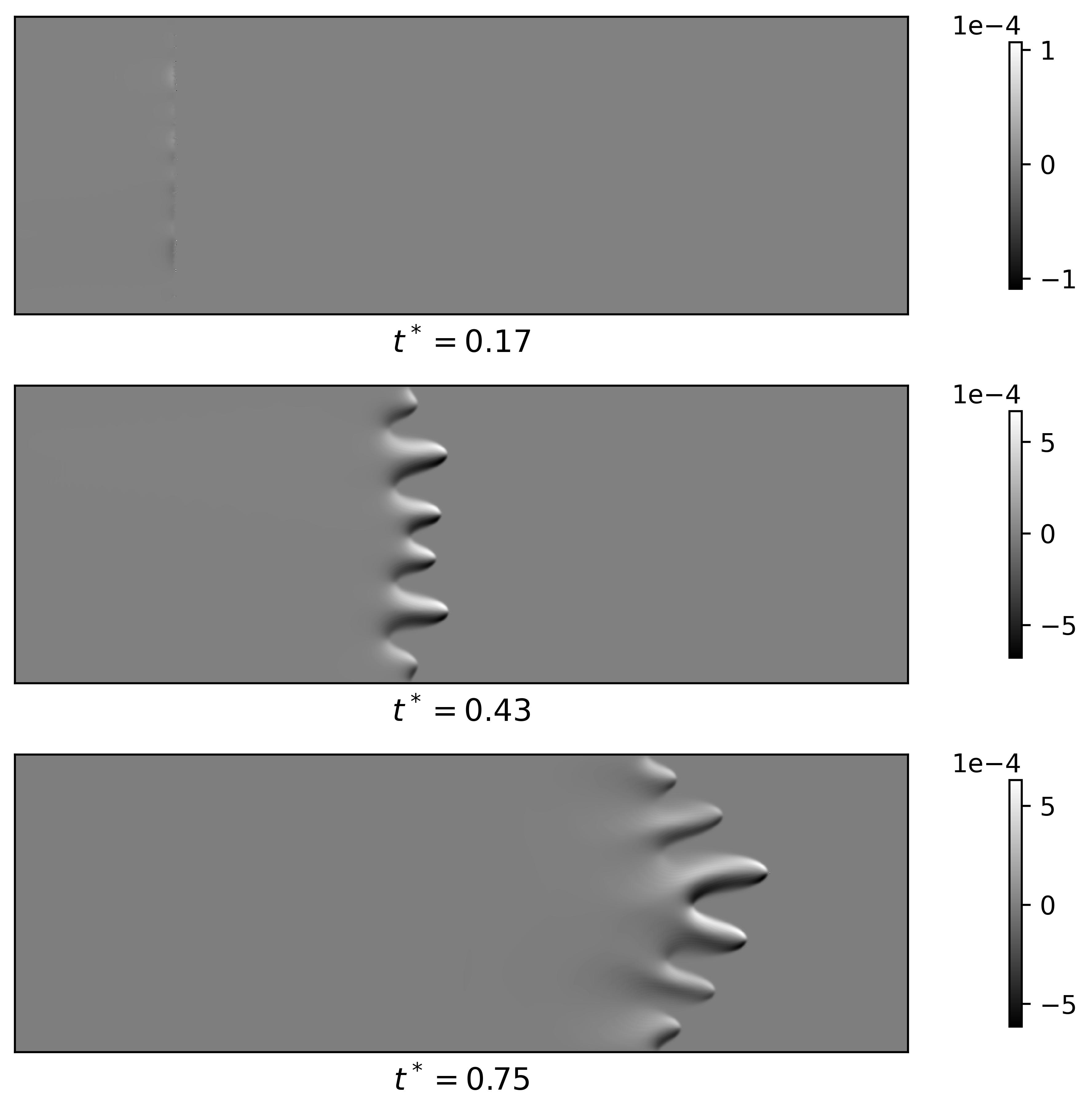}
    }\hfill
    \subfloat[$\nu{=}0.3\,\text{cm}^{-1}$\label{fig:cossw_vort_maps_b}]{
        \includegraphics[width=0.48\textwidth]{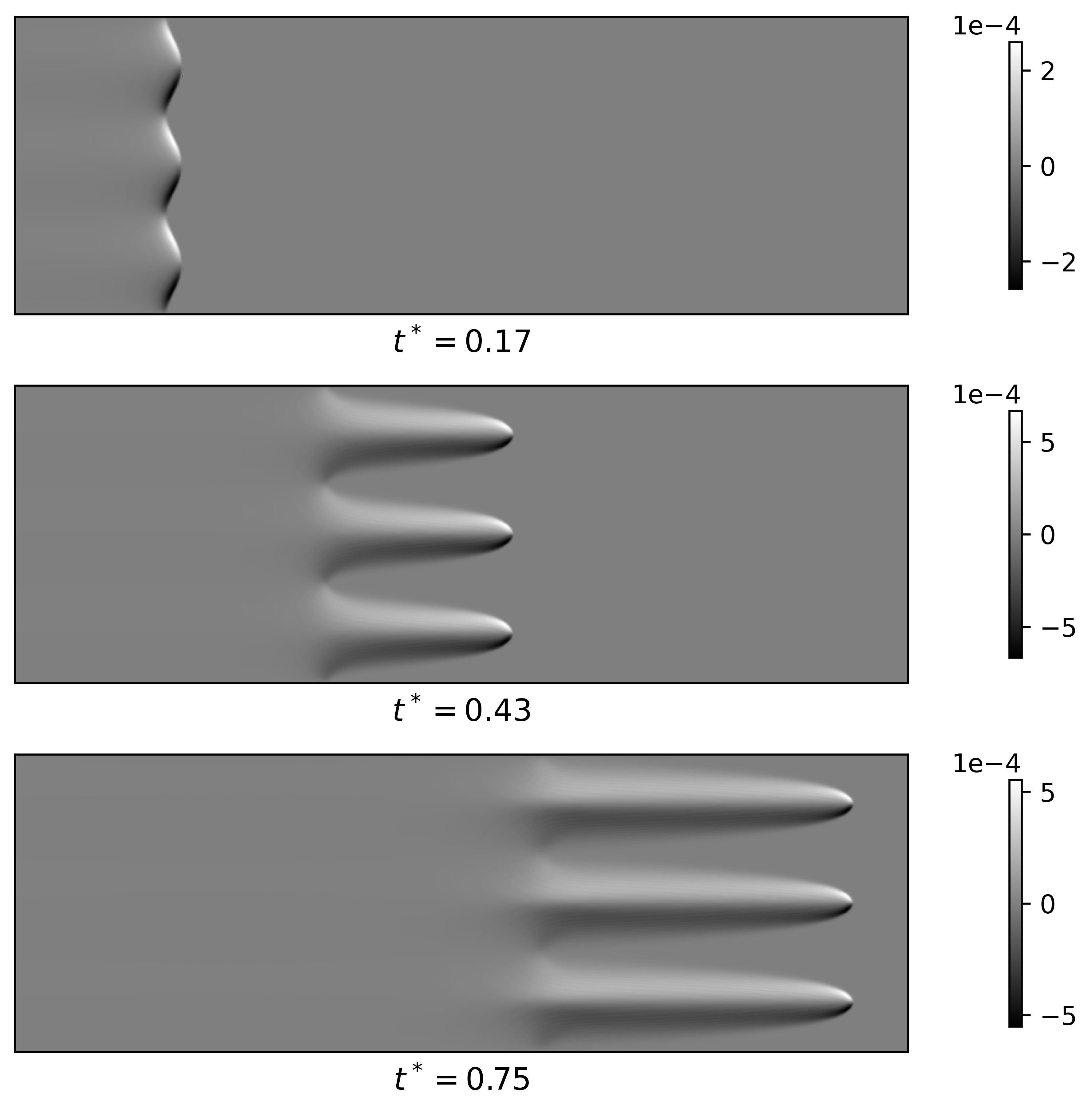}
    }\hfill
    \subfloat[$\nu{=}0.6\,\text{cm}^{-1}$\label{fig:cossw_vort_maps_c}]{
        \includegraphics[width=0.48\textwidth]{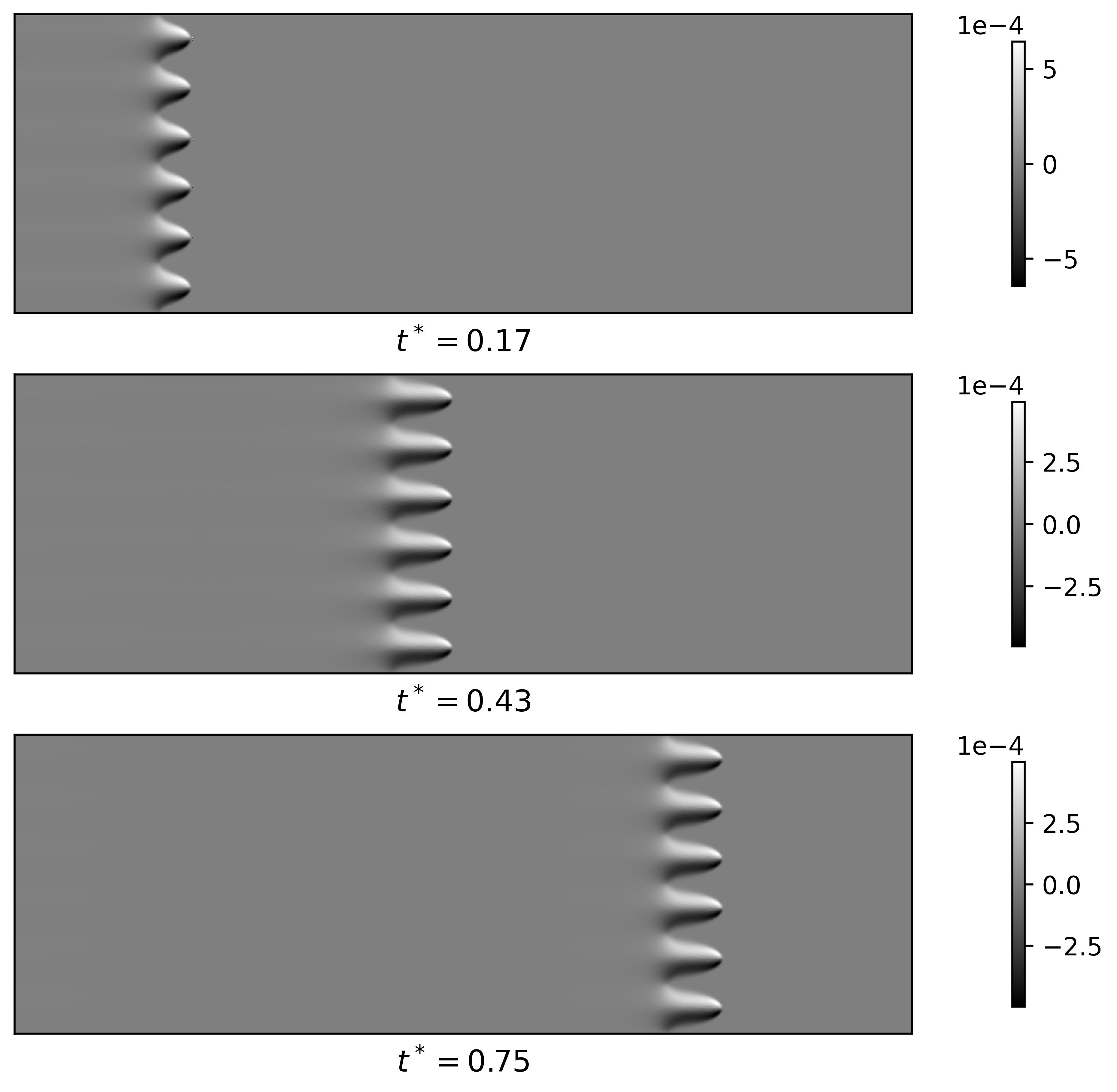}
    }\hfill
    \subfloat[$\nu{=}1.2\,\text{cm}^{-1}$\label{fig:cossw_vort_maps_d}]{
        \includegraphics[width=0.48\textwidth]{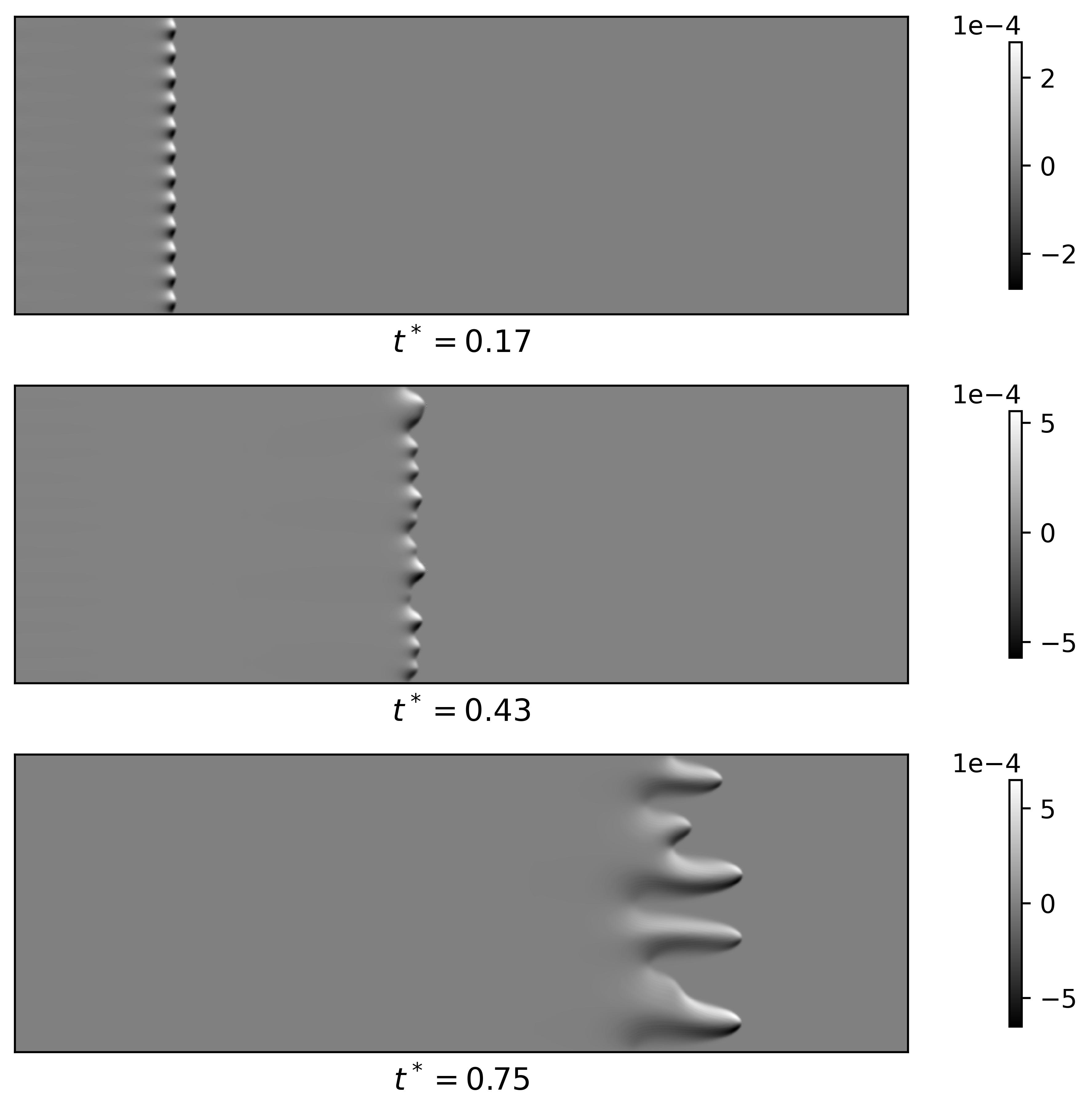}
    }\hfill
    \caption{Vorticity maps for the cases with $\nu{=}$0.1, 0.3, 0.6 and 1.2 cm$^{-1}$, at different times during the simulation.}\label{fig:cossw_vort_maps}
\end{figure}

In Figure~\ref{fig:cossw_size_sigma_krv6a_2}, we see that the growth rates and finger size curves display erratic behaviour for the low and high wavenumbers cases. This is driven by finger splitting, in the case of low wavenumbers, and by finger merging in the case of high wavenumbers. To better visualize these effects, as well the behaviour of the intermediate wavenumbers, we show, in Figure~\ref{fig:cossw_vort_maps}, the vorticity maps for the cases with $\nu{=}$0.1, 0.3, 0.6 and 1.2 cm$^{-1}$, at different times during the simulation.

If we compare the fingers of the ${\nu{=}0.3\,\text{cm}^{-1}}$ and ${\nu{=}0.6\,\text{cm}^{-1}}$ cases, we see that the thicker fingers (${\nu{=}0.3\,\text{cm}^{-1}}$), which initially grow slower than those from ${\nu{=}0.6\,\text{cm}^{-1}}$, maintain their growth and eventually become larger than those from ${\nu{=}0.6\,\text{cm}^{-1}}$, which stagnate in size after $t^*{=}0.3$.

\begin{figure}[!ht]
    \centering
    \includegraphics[width=0.48\textwidth]{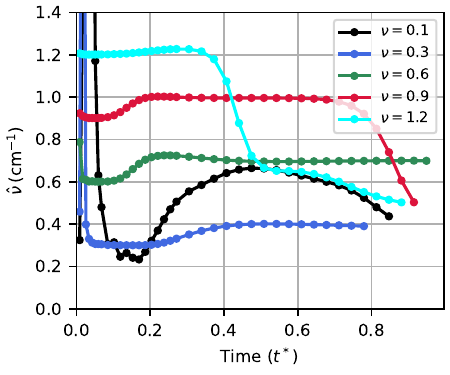}
    \caption{Evolution of the dominant wavenumber of the fingers over time. Each curve is shown up to the corresponding breakthrough time.}\label{fig:dominant_mode_2}
\end{figure}

For the case with ${\nu{=}1.2\,\text{cm}^{-1}}$, the fingers stagnate at much smaller values than those seen in the case with ${\nu{=}0.6\,\text{cm}^{-1}}$. That means the accumulation of numerical errors during the simulation is enough to distort the displacement front, causing some fingers to grow larger than their neighbours ($t^*{=}0.4$), which leads to shielding effects, where the larger fingers impede the growth of their smaller neighbours. As the simulation progresses, the smaller fingers eventually merge into the larger ones. This behaviour is consistent with the findings of \citet{riaz2006b} and can be seen in Figure~\ref{fig:dominant_mode_2}, which tracks the evolution of the dominant modes ($\hat{\nu}$) for cases with different initial perturbations. We see that for the ${\nu{=}1.2\,\text{cm}^{-1}}$ case, the dominant mode suddenly decreases at around ${t^*{=}0.4}$, indicating the merging of the fingers. The same is seen for the ${\nu{=}0.9\,\text{cm}^{-1}}$ case, at around ${t^*{=}0.8}$. As can be seen in Figures~\ref{fig:cossw_size_sigma_krv6a_2} and~\ref{fig:dominant_mode_2}, this collapse in $\hat{\nu}$ does not happen for cases with initial perturbations with ${\nu\leq0.8\,\text{cm}^{-1}}$, because the fingers generated from these perturbations are large enough to be unaffected by numerical errors (at least before breakthrough time).

The case with ${\nu{=}0.1\,\text{cm}^{-1}}$, which should display only one finger in the simulation, displays, instead, six fingers---which corresponds to the fastest growing wavenumber at the onset of instability, ${\nu{=}0.6\,\text{cm}^{-1}}$. This is a clear example of tip splitting where small numerical errors seed new fingers that grow faster than the original one, and explains the erratic behaviour of the growth rate for low wavenumbers in Figure~\ref{fig:cossw_sigma_krv6a_2}. Because the fingers generated are not of equal size, the larger ones eventually ``shield'' the smaller ones, which leads to a decrease in the dominant wavenumber of the fingers as the simulation progresses. This can be observed in Figure~\ref{fig:dominant_mode_2}, where we see that the calculated $\hat{\nu}$ of the ${\nu{=}0.1\,\text{cm}^{-1}}$ case starts to decline after it reaches a maximum of $\hat{\nu}\approx0.6\,\text{cm}^{-1}$, at around ${t^*{=}0.5}$. 

\begin{figure}[!ht]
    \centering
    \includegraphics[width=0.5\textwidth]{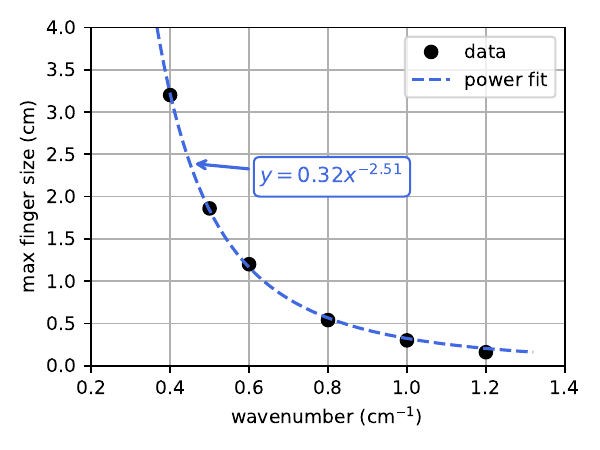}
    \caption{Maximum finger size as a function of initial wavenumber. Simulation results and the fitted power-law relation.}\label{fig:maxsize}
\end{figure}

In Figure~\ref{fig:maxsize}, we plot the maximum size of the fingers (before numerical instability provokes merging) as a function of the initial wavenumber, and we can see a power-law relation between the maximum size and the wavenumber (with an exponent of approximately 2.5), showing that the thicker (low wavenumber) fingers can grow substantially larger than the thinner (large wavenumber) ones. That fingers growth ceases in the nonlinear regime agrees with the weakly nonlinear analysis in \citet{chikhliwala1988b}. Although their analysis was limited to wavenumbers near the cut-off wavenumber, they find that perturbations with an unstable wavenumber eventually reach an equilibrium amplitude ($A_{eq}$) in the nonlinear regime, even if the initial perturbation amplitude was larger than $A_{eq}$. Our simulations show that this seems to be true for all unstable wavenumbers (though our grid dimensions limit the minimum wavenumbers we can check). The physical mechanism that arrests growth is relatively straightforward. As is known from LSA, transverse diffusion scales with the square of the wavenumber, which is what eventually leads to a cut-off wavenumber ($\nu_{cut}$) that does not destabilize the displacement. Although at the onset of the instability, such diffusive flow is not enough to arrest finger growth for cases where $\nu<\nu_{cut}$, as the fingers elongate (keeping their width) and the displacement front deforms, the total area (or perimeter) of the front increases, increasing the total transverse diffusive flow, at some point stabilizing the fingers. Although the fact that our power-law fit gives an exponent of 2.5, not 2, suggests some other nonlinear/geometrical effects are in play.

The arrest of finger growth in the nonlinear regime is conditional on the fingers maintaining their uniformity during the simulation, although the fact that transverse diffusion (the main stabilizing mechanism) is increased by the elongation of the fingers is not limited to the uniform fingers' scenario. In fact, as can be seen in~\cite{riaz2006b}, when fingers are randomly seeded, the longer fingers not only suppress the growth of their smaller neighbours by shielding effects, but also grow in thickness as they grow in length, eventually absorbing the smaller fingers. In summary, the widening of the fingers (causing merging) is linked with their ability to grow longer. 

Another interesting observation about the calculated $\hat{\nu}$ for the $\nu{=}$0.3, 0.6 and 0.9 cm$^{-1}$ cases (Figure~\ref{fig:dominant_mode_2}), is the gradual increase in $\hat{\nu}$, with an eventual stabilization at a new, higher value than the original perturbation wavenumber. This is because, as the fingers grow, their shapes start to deviate from the original cosine shape. This increases the value for $\hat{\nu}$ calculated from Equation~\ref{eq:dominant_mode}, even if the number of fingers remain unchanged.

\begin{figure}[!ht]
    \centering
    \includegraphics[width=\textwidth]{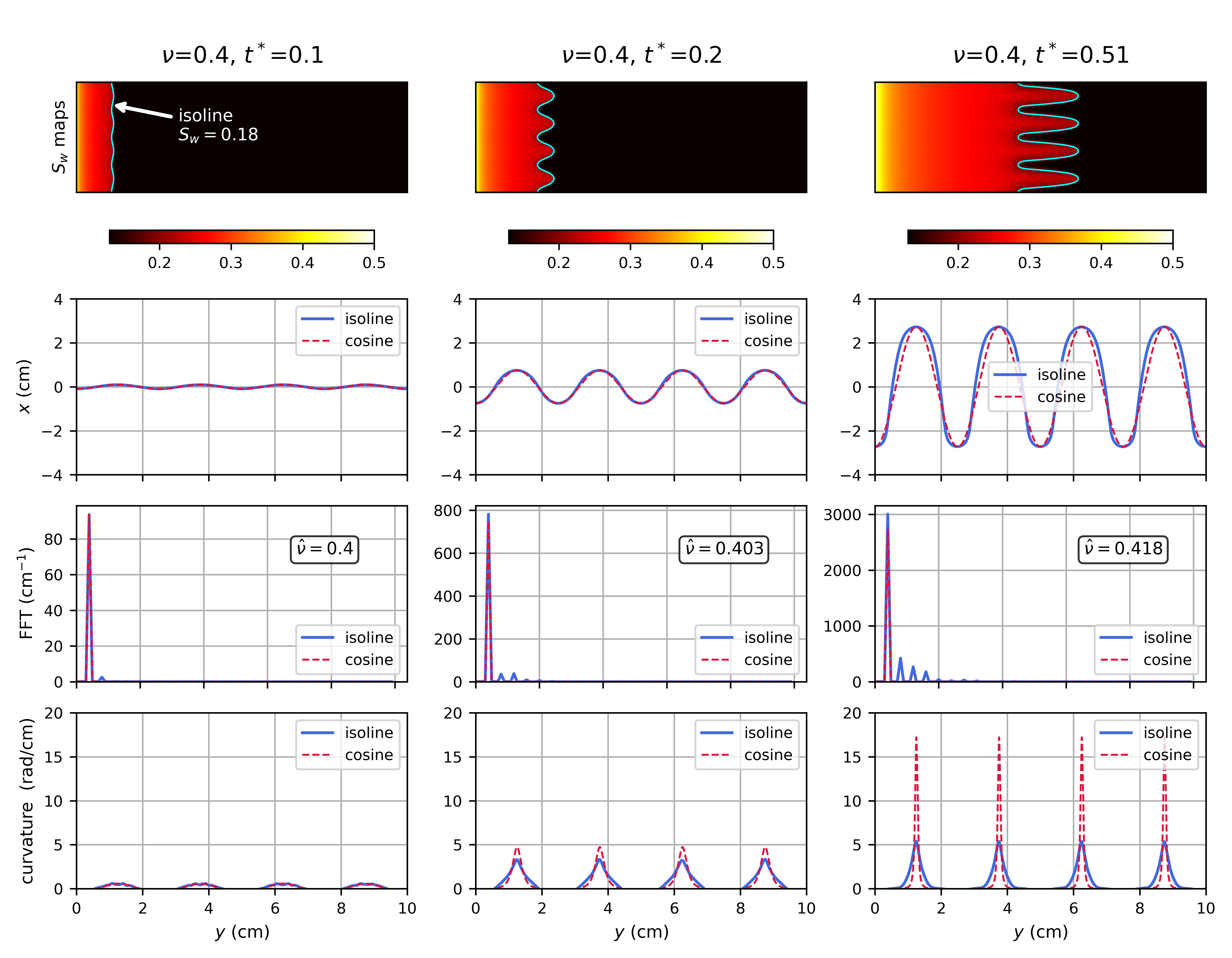}
    \caption{Saturation, finger profile and curvature for the ${\nu{=}0.4\,\text{cm}^{-1}}$ case, at two different times during the simulation. For each time, the top graph shows the $S_w$ maps and an isoline at $S_w{=}0.18$ (shown in blue). The second (from the top) graph compares the isoline shape with the cosine function. The third graphs compare the FFT of the isoline and cosine function. And the last graph shows the positive values of the curvature along the isoline.}\label{fig:fingers_isoline}
\end{figure}

We exemplify this in Figure~\ref{fig:fingers_isoline}, which shows, for the case with ${\nu{=}0.3\,\text{cm}^{-1}}$, at ${t^*=0.1}$ and ${t^*=0.51}$: the saturation distribution; the isoline $S_w{=}0.18$ (midway between $S_{w0}=0.13$ and $S_{ws}=0.24$) that defines the profile of the fingers/displacement front; the Fast Fourier Transform (FFT) of the isoline; and the curvature $\kappa$, given by:
\begin{equation}
    \kappa = \frac{\mathrm{d}^2 x}{\mathrm{d}y^2} \frac{1}{{\left( 1 + {\left( \mathrm{d}x / \mathrm{d}y \right)}^2 \right)}^{3/2}}.   
\end{equation}

\begin{figure}[!ht]
    \centering
    \includegraphics[width=\textwidth]{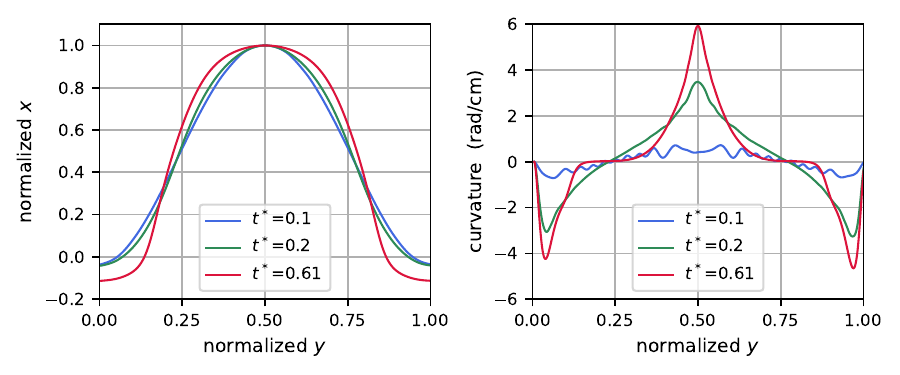}
    \caption{Normalized dimensions and curvature of one finger of the $\nu{=}0.4\,\text{cm}^{-1}$ case, for three different times during the simulation.}\label{fig:curvature_timevar}
\end{figure}

We see that as the fingers grow, the profile of the fingers (the isoline) deviates from a cosine-type curve, which leads to higher modes appearing in the Fourier transform of this profile, increasing the average value of the dominant mode, and accounts for the observed behaviour in Figure~\ref{fig:dominant_mode_2}. The fingers also become ``sharper'' as they grow longer (but retain their width), which is reflected in the increasing curvature. The tips of the fingers become more rounded as they grow, which can be seen more clearly in Figure~\ref{fig:curvature_timevar}, where we plot the normalized dimensions and curvature (not normalized) of one finger of the $\nu{=}0.4\,\text{cm}^{-1}$ case, for three different times.

To understand the impact of a curvature at the front, we change to a coordinate system that follows the curved front of the fingers, 
\begin{equation}
    \bol{r}(s, \eta, t) = \bol{r}_f(s, t) + \eta \hat{\bol{n}}(s, t),    
\end{equation}
where $\bol{r}_f(s, t)$ is the position of an isoline $S_w{=}S_{wf}$ at the displacement front, $s$ is the curvilinear coordinate along the isoline, $\hat{\bol{n}}(s, t)$ is the unit vector normal (pointing outwards) to the isoline at $s$, and $\eta$ is the signed normal distance. The Frenet-Serret relations gives us:
\begin{equation}
    \frac{\mathrm{d} \hat{\bol{t}}}{\mathrm{d} s} = -\kappa \hat{\bol{n}},\quad \frac{\mathrm{d} \hat{\bol{n}}}{\mathrm{d} s} = \kappa \hat{\bol{t}},  
\end{equation}
where $\hat{\bol{t}}(s,t){=}\mathrm{d}\bol{r}_f/\mathrm{d}s$ is the unit vector tangent to the isoline, $\kappa{=}\kappa(s, t)$ is the curvature of the isoline. Assuming that the transition zone at the front is small compared to the radius of curvature, i.e. $\eta / R = \eta \kappa \ll 1$, and that $\partial_{s} S_w$ and $\partial_{ss} S_w$ are small near $\eta{=}0$, we can write the diffusive term in the advection-diffusion equation as:
\begin{equation}
    \nabla \cdot \left( D_c \nabla S_w \right) \approx \frac{\partial}{\partial \eta} \left( D_c \frac{\partial S_w}{\partial \eta} \right) + \kappa D_c \frac{\partial S_w}{\partial \eta},
\end{equation}
where $D_c$ is the dispersive coefficient given by Equation~\ref{eq:Dc}.

The ${\kappa D_c \partial (S_w / \partial\eta)}$ of the above equation gives us the impact of the curvature. Given that $D_c<0$, at the tip of the fingers, where $\kappa>0$, this term acts as a convective term on the \textit{opposite} direction of flow. Conversely, at the region between the fingers, where $\kappa<0$, the curvature adds a convective term in the \textit{same} direction of flow. In both cases, the curvature has a stabilizing effect. This helps explain why the tip of the fingers become more rounded as they grow, as seen in Figure~\ref{fig:fingers_isoline}. As can be seen by the comparison between the isoline and the cosine function curvatures (bottom-right graph), the curvature at the tip of the cosine function is much greater than at the tip of the isoline. This peak of curvature at the tip would mean a concentrated stabilizing flow at the tip, eventually ``rounding'' the shape of the fingers.

\begin{figure}[!ht]
    \centering
    \includegraphics[width=\textwidth]{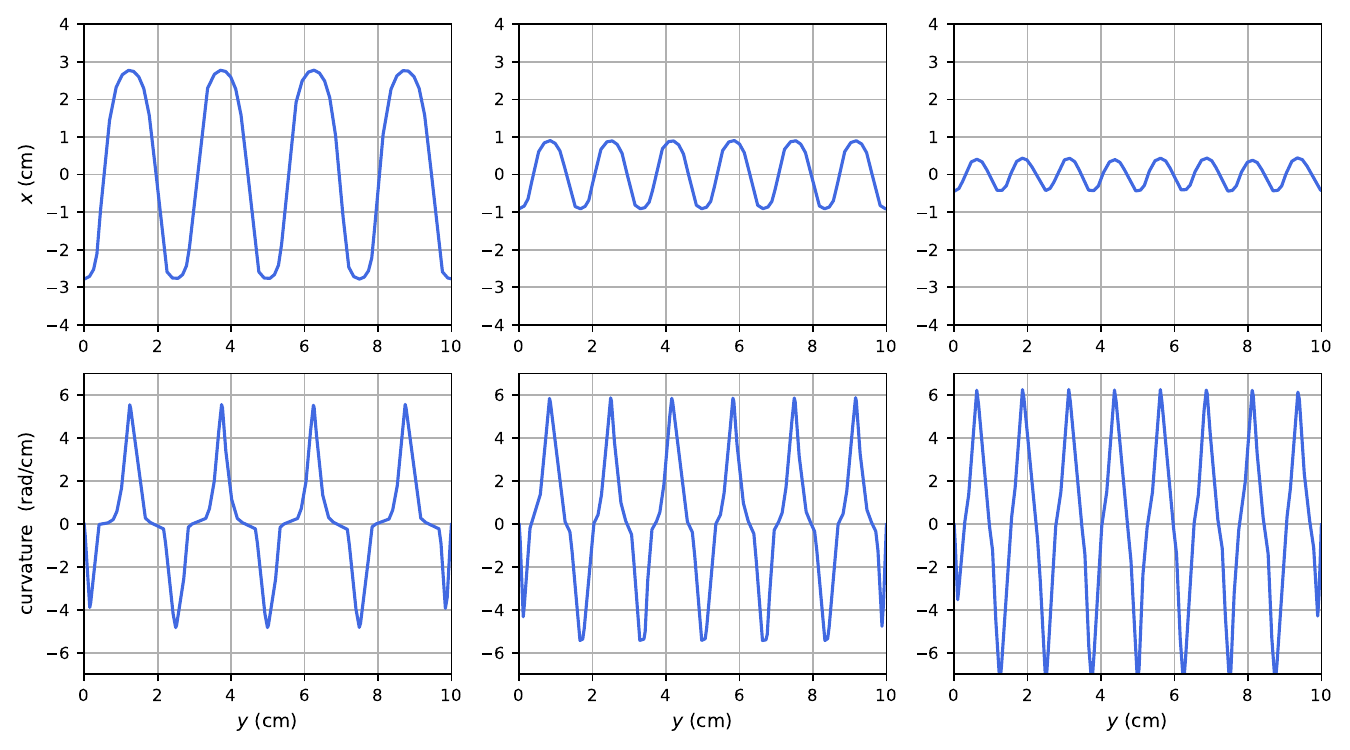}
    \caption{Finger profile and curvature for the cases with $\nu{=}$0.4, 0.6 and 0.8 cm$^{-1}$, after finger growth has ceased.}\label{fig:curvature_casevar}
\end{figure}

\begin{figure}[!ht]
    \centering
    \includegraphics[width=\textwidth]{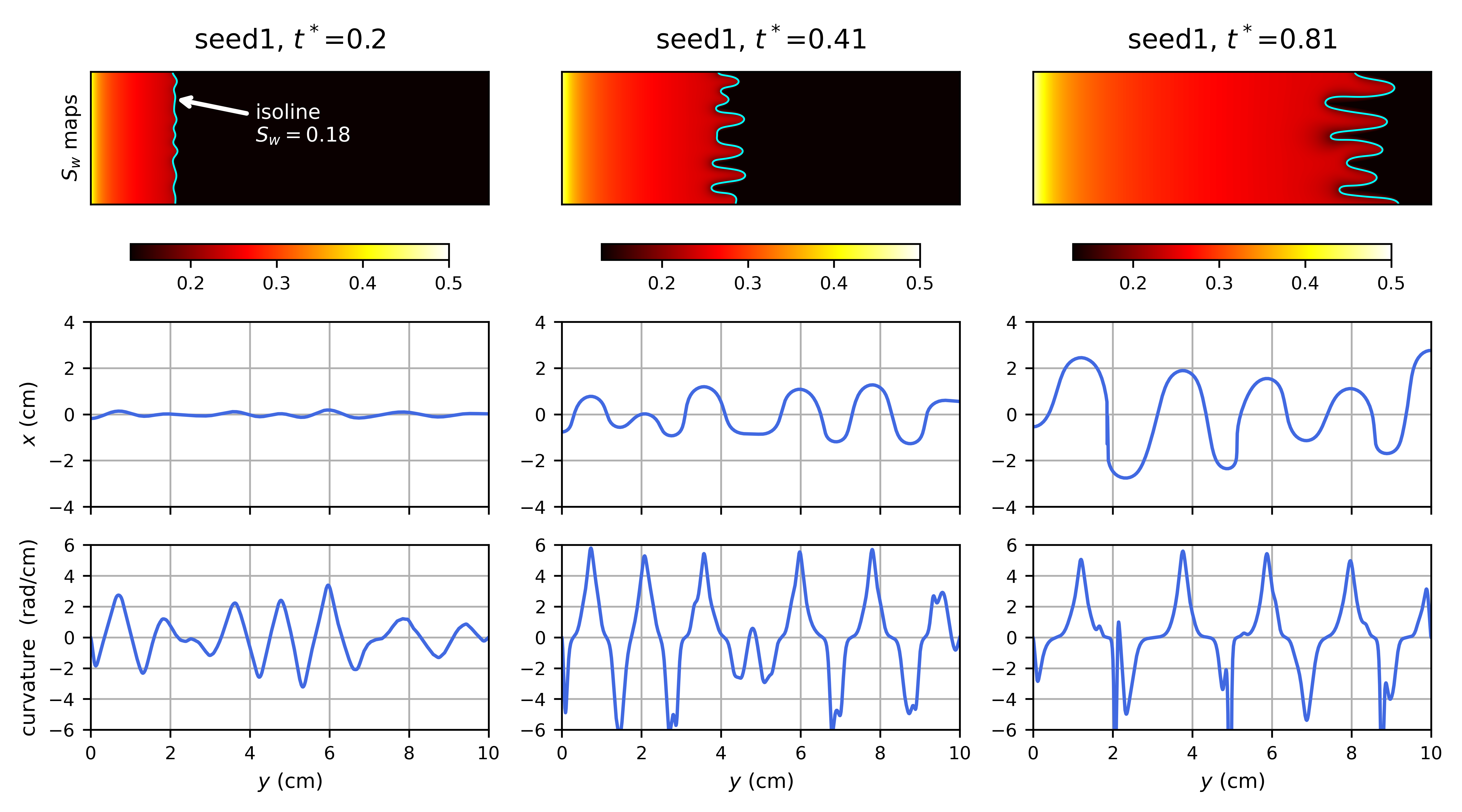}
    \caption{Saturation, finger profile and curvature for the case with random perturbations at the inlet. For each time, the top graph shows the $S_w$ maps and an isoline at $S_w{=}0.18$ (shown in blue). The second (from the top) graph compares the isoline shape with the cosine function. And the last graph shows the curvature along the isoline.}\label{fig:curvature_random}
\end{figure}

Interestingly, Figure~\ref{fig:curvature_casevar} shows that though the dimensions of the stagnated fingers differ considerably for the cases with $\nu{=}$0.4, 0.6 and 0.8 cm$^{-1}$, their curvatures at their tips are similar, which suggests a local balance is reached at the tips of the fingers, and that this balance is independent, or weakly dependent, of the wavelength (or finger thickness). Further support for this hypothesis of local balance comes from analysing the curvature of fingers seeded from random perturbations, shown in Figure~\ref{fig:curvature_random}. We can observe that, though the fingers keep growing and merging, the maximum curvature values at the tip remain relatively constant after some time, and very similar to the maximum curvature values observed for the wavelike perturbations.

%%%% References %%%%
\FloatBarrier% chktex 1
\bibliography{references.bib}% common bib file